%% file: Tremblay12_Paper2_revision1.tex
\documentclass[usenatbib]{mn2e}

\pdfoutput=1

\usepackage{natbib}
\usepackage{graphicx}
\usepackage{apjfonts}
\usepackage{deluxetable}
\usepackage{multicol}
\usepackage{grant_defs}
\usepackage{verbatim}

\usepackage{aastex_hack}

\title[AGN Feedback in Abell 2597]{Multiphase Signatures of AGN Feedback in Abell 2597}

\author[G.~R.~Tremblay et al.]{G.~R.~Tremblay,$^{1,2,3}$ C.~P.~O'Dea,$^{2,4}$ S.~A.~Baum,$^{3,5}$ T.~E.~Clarke,$^{6}$ C.~L.~Sarazin,$^{7}$
\newauthor J.~N.~Bregman,$^{8}$ F.~Combes,$^{9}$ M.~Donahue,$^{10}$ A.~C.~Edge,$^{11}$ A.~C.~Fabian,$^{12}$ G.~J.~Ferland,$^{13}$
\newauthor  B.~R.~McNamara,$^{4,14}$ R.~Mittal,$^{3}$ J.~B.~R.~Oonk$^{15}$ A.~C.~Quillen,$^{16}$ H.~R.~Russell,$^{14}$ 
\newauthor J.~S.~Sanders,$^{12}$ P.~Salom\'{e},$^{9}$ G.~M.~Voit,$^{10}$ R.~J.~Wilman,$^{11}$ and M.~W.~Wise$^{15}$ 
\\\\$^{1}$ European Southern Observatory, 
Karl-Schwarzschild-Str.~2, 85748 Garching bei M\"{u}nchen, Germany; grant.tremblay@eso.org 
\\$^{2}$ Department of Physics, Rochester Institute
of Technology, 84 Lomb Memorial Drive, Rochester, NY 14623, USA
\\$^{3}$ Chester F.~Carlson Center for Imaging Science,  54 Lomb Memorial Drive, Rochester, NY 14623, USA
\\$^{4}$ Harvard-Smithsonian Center for Astrophysics, 60 Garden St., Cambridge, MA 02138, USA
\\$^{5}$ Radcliffe Institute for Advanced Study, 10 Garden St., Cambridge, MA 02138, USA
\\$^{6}$ Naval Research Laboratory Remote Sensing Division, Code 7213 4555 Overlook Ave SW, Washington, DC 20375, USA
\\$^{7}$ Department of Astronomy, University of Virginia, P.O. Box 400325, Charlottesville, VA 22904-4325, USA
\\$^{8}$ University of Michigan, Department of Astronomy, Ann Arbor, MI 48109, USA
\\$^{9}$ Observatoire de Paris, LERMA, CNRS, 61 Av.~de l'Observatoire, 75014 Paris, France
\\$^{10}$ Michigan State University, Physics and Astronomy Dept., East Lansing, MI 48824-2320, USA
\\$^{11}$ Department of Physics, Durham University, Durham, DH1 3LE, UK
\\$^{12}$ Institute of Astronomy, Madingley Rd., Cambridge, CB3 0HA, UK
\\$^{13}$ Department of Physics, University of Kentucky, Lexington, KY 40506, USA
\\$^{14}$ Physics \& Astronomy Dept., Waterloo University, 200 University Ave.~W., Waterloo, ON, N2L, 2G1, Canada
\\$^{15}$ ASTRON, Netherlands Institute for Radio Astronomy, P.O. Box 2, 7990 AA Dwingeloo, The Netherlands
\\$^{16}$ Department of Physics and Astronomy, University of Rochester, Rochester, NY 14627, USA}

\begin{document}

\input{journals}

\date{Accepted for Publication in MNRAS, 9 May 2012}

\pagerange{\pageref{firstpage}--\pageref{lastpage}} \pubyear{2012}

\maketitle

\label{firstpage}

\begin{abstract}
We present new {\it Chandra} X-ray observations of the brightest cluster galaxy (BCG) in the cool core cluster Abell 2597 ($z=0.0821$). 
The data reveal an extensive kpc-scale X-ray cavity network as well as a 15 kpc filament of soft-excess gas exhibiting strong spatial correlation with archival VLA radio data. 
In addition to several possible scenarios, multiwavelength evidence may suggest that the filament is associated with multiphase ($10^{3}-10^{7}$ K) gas that has been entrained and dredged-up by the propagating radio source.
Stemming from a full spectral analysis, we also present profiles and 2D spectral maps of modeled X-ray temperature, 
entropy, pressure, and metal abundance. The maps reveal 
an arc of hot gas
which in  projection
borders the  inner edge  of a large  X-ray cavity. 
Although limited by strong caveats, we suggest that the hot arc may be (a) due to a compressed rim of cold gas pushed outward by the radio bubble or (b)   morphologically and energetically consistent with cavity-driven active galactic nucleus (AGN) heating models invoked to quench cooling flows, in which 
the enthalpy of a buoyant X-ray cavity is locally thermalized as ambient gas rushes to refill its wake. If confirmed, this would be the first observational evidence for this model. 
\end{abstract}


\section{Introduction}
\label{section:introduction}

Although an important component  in models of galaxy evolution, little
is  known about  the  mechanical interaction  between active  galactic
nucleus (AGN)  outflows and  the ambient gaseous  environments through
which  they  propagate.  Observations  of brightest  cluster  galaxies
(BCGs) in cool  core (CC) clusters suggest that  the outflowing plasma
can  drive  sound waves  and  subsonically  excavate  cavities in  the
ambient    X-ray    bright    intracluster    medium    (ICM,    e.g.,
\citealt{sarazin86}),  acting as lower-limit  calorimeters to  the AGN
kinetic                          energy                          input
\citep{boehringer93,fabian00,fabian06,churazov01,mcnamara00,mcnamara01,blanton01,nulsen05,forman05,forman07,birzan04}.
Along  with  a wealth  of  supporting  circumstantial evidence,  these
observations motivate the radio-mode AGN feedback paradigm, invoked at
late  epochs  to  inhibit   catastrophic  cooling  flows  that  should
otherwise drive extreme star formation rates in the BCG (e.g., reviews
by  \citealt{fabian94,peterson06,mcnamara07,mcnamara12,fabian12}).    However,  while  the
cavities demonstrate the profound impact  of jets on ambient hot X-ray
haloes, the  physics coupling AGN  mechanical energy to  ICM structure
and entropy remain poorly understood.

Moreover, models  invoking radio-mode feedback to  quench cooling flow
signatures   such  as   star   formation  must   be  reconciled   with
observational evidence that, in several systems, the propagating radio
source triggers rather than inhibits star formation. {\it Hubble Space
  Telsescope} ({\it  HST}) FUV observations of several  CC BCGs reveal
knots of young  stars preferentially aligned along the  edges of radio
lobes --- sites where the outflowing plasma collides with the observed
ambient  cold gas  (e.g.,  Abell 1795,  Abell 2597,  
\citealt{odea04,tremblay_cooling},  and  references therein).  Many studies  have
suggested that the jet propagation  front can shock and compress these
cold   clouds,  inducing   rapid,  short-duration   starbursts  (e.g.,
\citealt{elmegreen78,voit88,deyoung89,mcnamara93,odea04,batcheldor07,holt08,holt11}). Mass
loading of  the jet can  similarly induce deceleration and  bending of
the radio source  in regions of high gas  density, which could explain
the compact,  steep spectrum, and bent \citet{fanaroff74}  class I (FR
I)  morphologies commonly  associated with  radio sources  embedded in
dense environments (e.g., \citealt{chrisgreenbank,baum88}).

Progress in understanding these issues  rests on a better grasp of the
complex ways in  which AGN outflows interact  not only
with their hot  gaseous environments, but with the  colder ISM phases as
well. In  this paper we present new  {\it Chandra X-ray
  Observatory} observations,  totaling 150 ksec  in combined effective
exposure time, of the brightest  cluster galaxy (BCG) at the centre of
the cool  core cluster Abell  2597 ($z=0.0821$).  Previous  works have
shown that the  source exhibits extensive, multiphase ($10^{3}-10^{7}$
K)  signatures  of  AGN/ISM  interactions, including  X-ray  cavities,
evidence for gas entrainment,  and jet-triggered star formation (e.g.,
\citealt{sarazin95,taylor99,odea04,clarke05,oonk10}).  In Section 2 we
describe the  observations, data reduction, and our  procedure for the
creation of 2D spectral maps from the new X-ray data.  In Section 3 we
present spatial and spectral results, including general X-ray morphology,
radio/X-ray spatial correlations, surface brightness and spectral parameter radial profiles, 
and a hardness analysis.  In Section
4  we present the  2D spectral  maps of  modeled X-ray  gas properties
including  temperature, entropy,  pressure, and  metal  abundance.  In
Sections 5 and 6 we discuss the main results of this paper, namely the
``cold  filament''  and ``hot  arc''  features  (respectively).  Throughout  this  work, we  adopt $H_0  =
71~h_{71}^{-1}$  km  s$^{-1}$   Mpc$^{-1}$,  $\Omega_M  =  0.27$,  and
$\Omega_{\Lambda} =  0.73$.  In this  cosmology, 1\arcsec\ corresponds
to $\sim 1.5$ kpc at the redshift of the A2597 BCG ($z=0.0821$).  This
redshift corresponds  to an  angular size distance  of $D_A\approx315$
Mpc and a luminosity distance of $D_L\approx369$ Mpc.  A discussion of
these new  X-ray observations  in the context  of ICM cooling  and AGN
heating  can be  found in  a companion  paper \citep{tremblay_cooling}.

\section{Observations \& Data Reduction}
\label{section:observations}

A summary of all new and archival data used (directly or referentially) 
in this analysis can be found in Table 1 of \citet{tremblay_cooling}. 
Here we describe the two new {\it Chandra X-ray  Observatory} AXAF CCD
Imaging Spectrometer (ACIS)  exposures of the A2597 BCG,  taken in May
2006 for a total of 52.8  and 60.9 ksec, respectively (ObsIDs 6934 and
7329, PI: Clarke). Both observations were taken in VFAINT, full-frame,
timed exposure mode.   An older 40 ksec {\it  Chandra} exposure (ObsID
922, PI: McNamara) was taken  in July 2000 and originally published by
\citet{mcnamara01}  (hereafter M01)  in  a study  of  the A2597  ghost
cavity (which  we will discuss  later). This observation was  taken in
FAINT, full-frame, timed exposure mode. As noted in M01, ObsID 922 was
badly impacted by flaring events,  yielding only $\sim18$ ksec of data
suitable  for spectral  analysis. We  have re-processed  all  of these
observations in a uniform manner, as described below.

The three observations  (ObsIDs 922, 6934, and 7329)  were obtained as
primary data products  from the {\it Chandra} Data  Archive.  The data
were  reduced in  version  4.2  of the  {\sc  ciao} environment  ({\it
  Chandra}       Interactive      Analysis       of      Observations,
\citealt{fruscione06}) using version 4.3.1 of the calibration database
(CALDB).  The  {\sc ciao}  script \texttt{chandra\_repro} was  run for
each observation to automate  the creation of new \texttt{level=1} and
\texttt{level=2} event files  by applying charge transfer inefficiency
(CTI)  correction,  time-dependent  gain  adjustment,  PHA  and  pixel
randomization,  and standard  grade and  status filtering.  Only those
events with  {\it ASCA}  grades 0,  2, 3, 4,  and 6  are used  in this
analysis. Background light curves for each exposure were obtained from
the   ACIS-S1   back-illuminated   chip.    The  {\sc   ciao}   script
\texttt{deflare} was used to  create good time interval (GTI) temporal
masks  rejecting counts associated  with three  sigma flares  over the
mean quiescent background rate.  This filtering strategy retained only
$\sim  18$  (out   of  40)  ksec  of  flare-free   data  for  the  922
observation. ObsIDs  6934 and 7329 were not  significantly impacted by
flaring events,  and no more than  2 ksec for  either observation were
rejected. The  combined effective exposure  time is 150 ksec,  and the
combined flare-free exposure time is 128 ksec.

A 0.5-7 keV energy filter was applied to the reduced exposures using standard
\texttt{dmcopy} techniques. The \texttt{fluximage} script was then used to create 
exposure maps for each dataset.
The observations were reprojected to a common WCS tangent point, combined, 
and divided by their similarly reprojected and combined exposure maps, creating exposure-corrected
0.5-7 keV mosaics suitable for spatial analysis. Soft (0.5-1 keV), medium (1-2 keV), and hard (2-7 keV) combined images were also created using this strategy. Adaptively smoothed 
and unsharp mask images were made using techniques discussed in Section 3. 
Merged data are not suitable for 
spectral analysis, so X-ray spectra were extracted from each individual observation 
and analyzed simultaneously in {\sc XSPEC} version 12.5 \citep{arnaud96}. 
We have also created 
2D spatial+spectral maps of projected X-ray temperature, pressure, and metal abundance, following a procedure described below.

\begin{figure*}
\begin{center}
\includegraphics[scale=0.50]{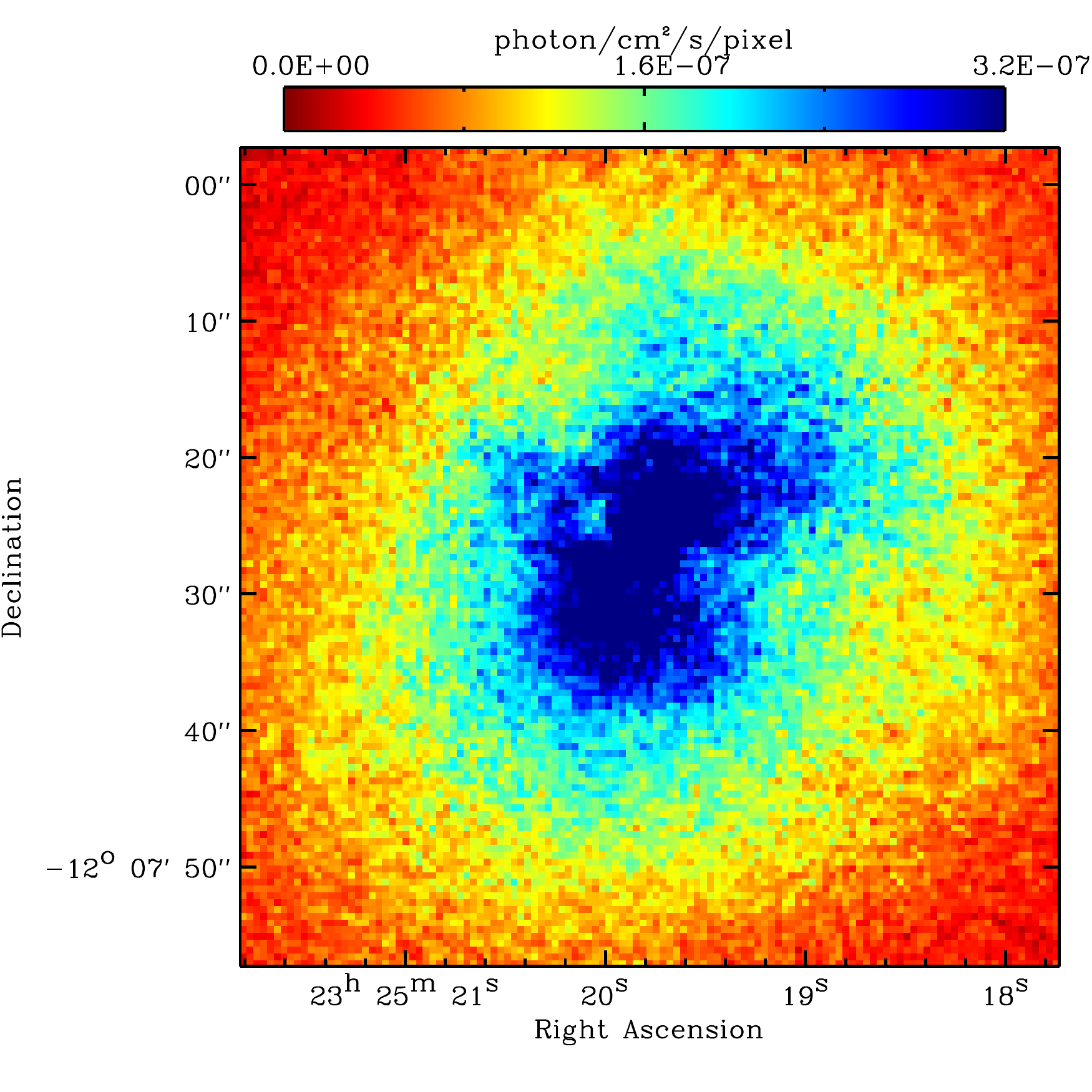}
\includegraphics[scale=0.48]{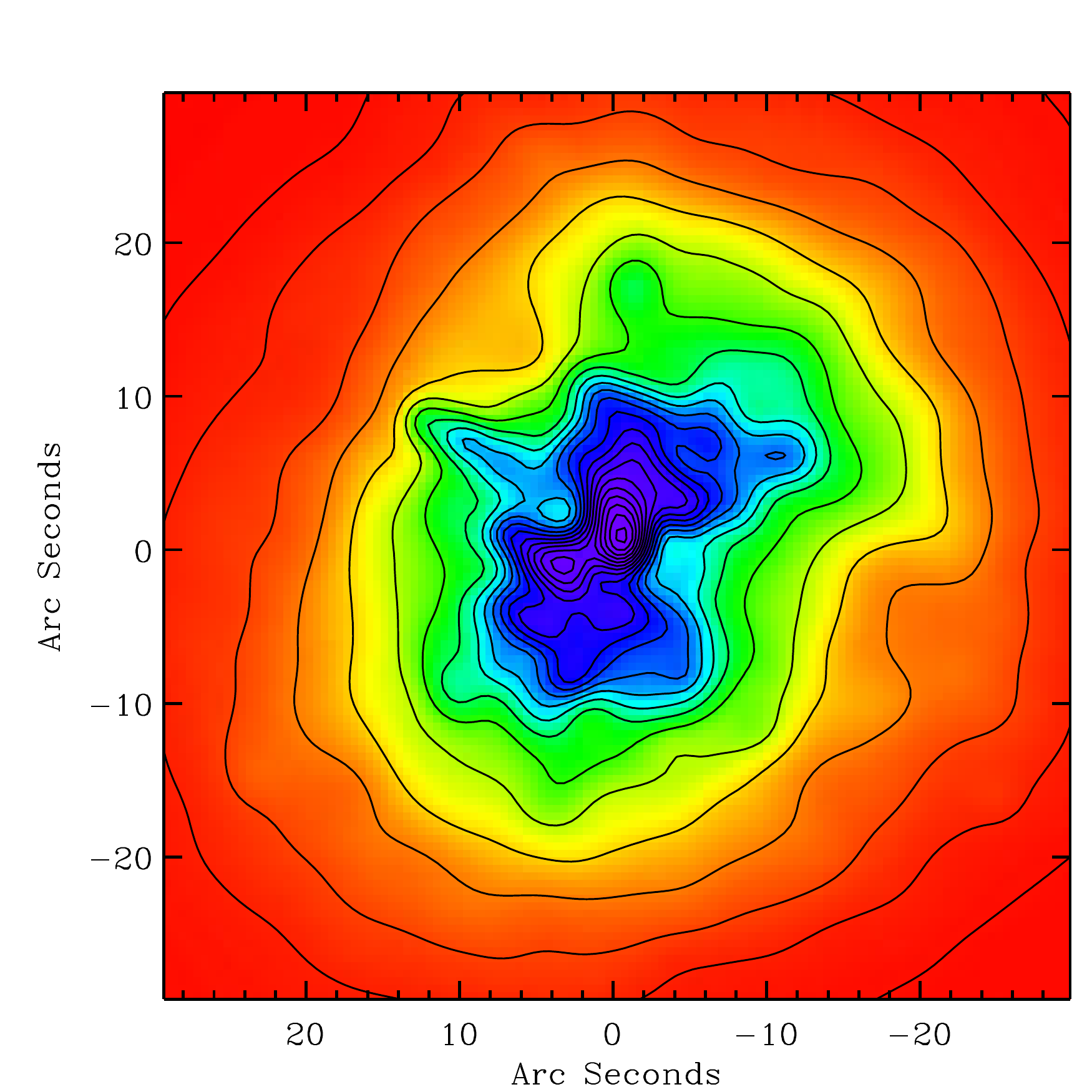}
\end{center}
\caption{({\it left})
Exposure corrected (fluxed) 0.5-7 keV image of the 
three merged {\it Chandra} exposures 922, 6934, and 7329.
No gaussian smoothing or pixel binning has been applied. 
({\it right}) 0.5-7 keV X-ray surface brightness map, smoothed with 
an adaptive gaussian kernel. 
Black  contours have  been overlaid  to
  better  show   the spatially anisotropic nature 
  of the emission. The innermost contour marks 
  a flux of $4.6\times10^{-7}$ photons sec\mone\ cm\mtwo\ pixel\mone, 
  and the contours move outward with a flux decrement of $2.0\times10^{-8}$
  photons sec\mone\ cm\mtwo\ pixel\mone. 
  Both panels share an identical field of view, centered at  RA=23h 25m 19.75s, Dec =-12$^\circ$ 07' 26.9'' (J2000). }
\label{fig:chandra}
\end{figure*}

\subsection{Creation of X-ray spectral maps}

Using the spatially resolved {\it Chandra} spectroscopy, we have created 2D 
maps of projected X-ray temperature, pseudo-entropy, pseudo-pressure, and metal abundance.
We briefly describe how these maps were created below, but refer the reader to Section 3 for more details on (e.g.) how spectra are extracted and modeled.

To make the  spectral maps, the exposure corrected  mosaic was cleaned
of  contaminating point  sources, cropped  to include  only  the inner
$80\arcsec\times80\arcsec$  ($120\times120$ kpc), then  passed through
the
\texttt{CONTBIN}\footnote{http://www-xray.ast.cam.ac.uk/papers/contbin/}
adaptive binning  algorithm described by  \citet{sanders05}.  The code
locates the  brightest pixel  in the image  and creates a  spatial bin
around  it  by  including  all  neighboring  pixels  of  like  surface
brightness until a user-defined  signal-to-noise threshold is met, and
iteratively moves outward following the same procedure.  For the X-ray
temperature and emission measure  maps (used in the pseudo-entropy and
pressure maps),  we set this threshold  to be S/N$=30$,  or $\gae 30^2
=900$ counts. For the metal abundance map, we set it to S/N$=70$.  The
shape of each spatial bin was  constrained so that its length could be
at most  two times  its width (to  prevent ``stripe''-like  bins which
would sample an unreasonably large spatial cross section of multiphase
cluster gas). The result is  a spatially binned image that follows the
surface brightness distribution.   In  addition  to the  contour  binned image,  the
algorithm  creates   a  bin  map   which  we  used   with  J.~Sanders'
\texttt{make\_region\_files}  code  to  create  {\sc  ciao}-compatible
region files for  each of the 277 individual  spatial bins created for
our   data    by   the   \texttt{CONTBIN}    code.    The   individual
\texttt{level=2} event  files for {\it  Chandra} ObsIDs 6934  and 7329
(which were  similarly cleaned of  point sources) were  then spatially
reprojected  to  identical  positions  using  the merged  image  as  a
coordinate reference. The  region files created from the  bin map were
then  ported to each  of the  two event  files. A  script was  used to
iteratively extract  the source spectrum, background  file, and create
associated  response  files (ARF/RMF)  for  each  of  the 277  spatial
regions on each of the  two event files.  The corresponding source and
background spectra  from like regions  on the two exposures  were then
added  together using  the \texttt{mathpha}  tool in  {\sc  ciao}, and
average   responses    were   created   using    \texttt{addarf}   and
\texttt{addrmf}  from  {\sc  ftools}.  The summed  spectra  were  then
regrouped   to  15   count   bins,  and   the  \texttt{BACKSCAL}   and
\texttt{EXPOSURE}  header keywords  were updated.   This  procedure is
only viable  if the exposures  were taken on  the same chip  (which is
true for  ObsIDs 6934 and 7329).  The 18 ksec (good  time) 922 dataset
was therefore not used in this part of the analysis.

An  XSPEC   v12.5  \citep{arnaud96}  TCL   script  was  used   to  fit
Mewe-Kaastra-Liedahl thermal  plasma models (called  \texttt{MEKAL} in
XSPEC),   absorbed  to   account  for   Galactic   attenuation  (i.e.,
\texttt{WABS}   $\times$  \texttt{MEKAL}).   These  models   were  fit
simultaneously  to  the  two  grouped,  background-subtracted  spectra
extracted from each of the  matched regions of the two exposures.  For
each fit,  the source redshift  was fixed to $z=0.0821$,  the hydrogen
column  density $N_\mathrm{H}$  was frozen  to the  Galactic  value of
$2.48  \times10^{20}$   cm$^{-2}$  \citep{mcnamara01},  and   the  gas
temperature  $kT$, abundance  $Z$, and  normalization $N_\mathrm{MEK}$
(effectively  the emission  measure)  were allowed  to  vary.  The
fits  were run  iteratively to  minimize the  $\chi^2$  statistic. The
best-fit   parameters   $kT$,   $Z$,   $N_\mathrm{MEK}$,   and   their
corresponding upper  and lower 90\% confidence  intervals were written
to  individual  files  linked  to  their  corresponding  spatial  bin.
J.~Sanders'  script \texttt{paint\_output\_images}  was  then used  to
rescale the spatial bins in the \texttt{CONTBIN} output image by their
associated  best-fit   parameters,  producing  projected  temperature,
emission measure, and abundance maps.  A projected pseudo-pressure map
was  then created  by  multiplying  the square  root  of the  emission
measure map  (which is  proportional to but  not exactly  the density,
which  requires an  uncertain  volume assumption)  by the  temperature
($kT$) map.   The maps  were then adaptively  smoothed using  the same
variable-width  Gaussian kernel  sizes used  for the  smoothed surface
brightness map shown in Fig.~\ref{fig:chandra}{\it b}.

\begin{figure*}
\begin{center}
\includegraphics[scale=0.50]{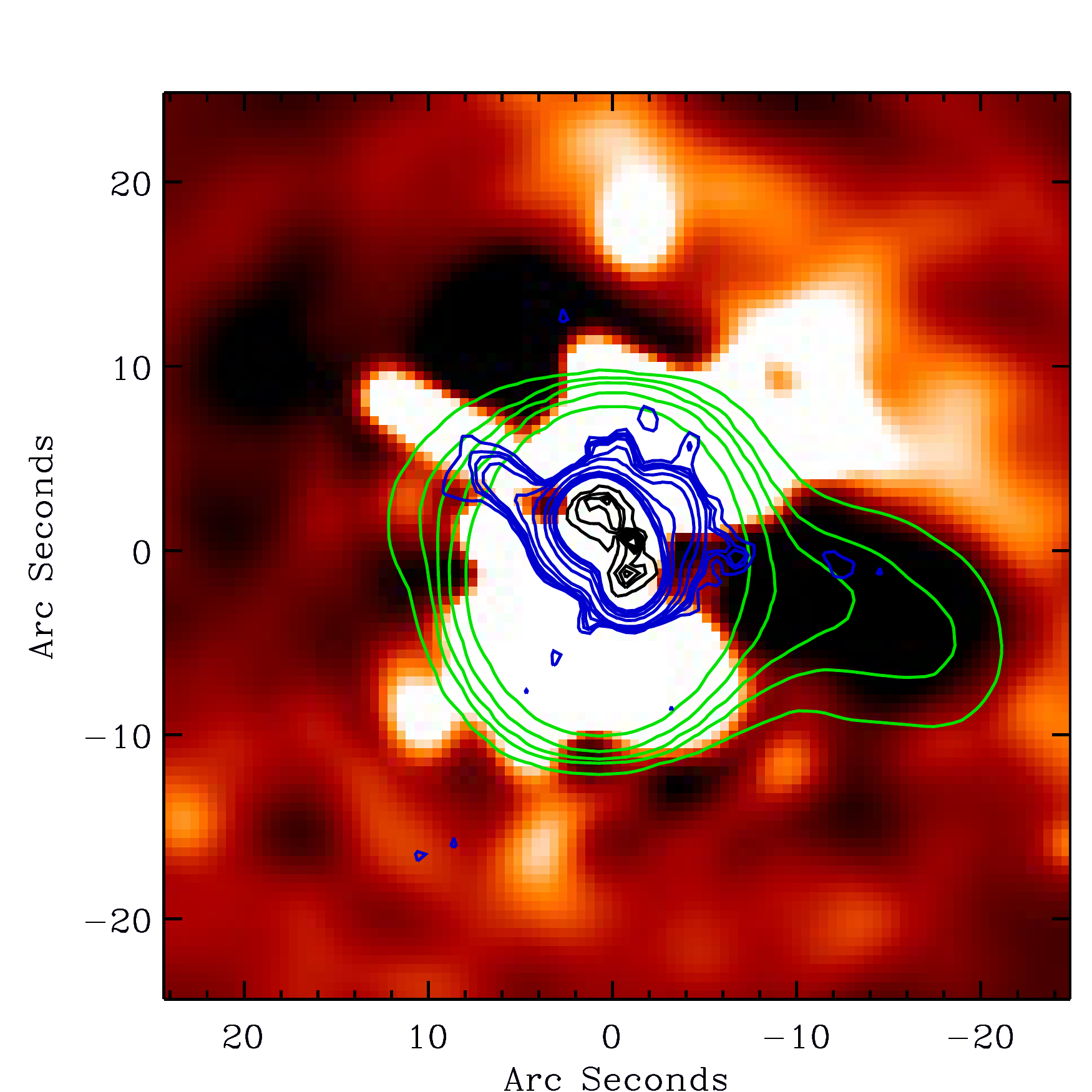}\hspace*{-2mm}
\includegraphics[scale=0.50]{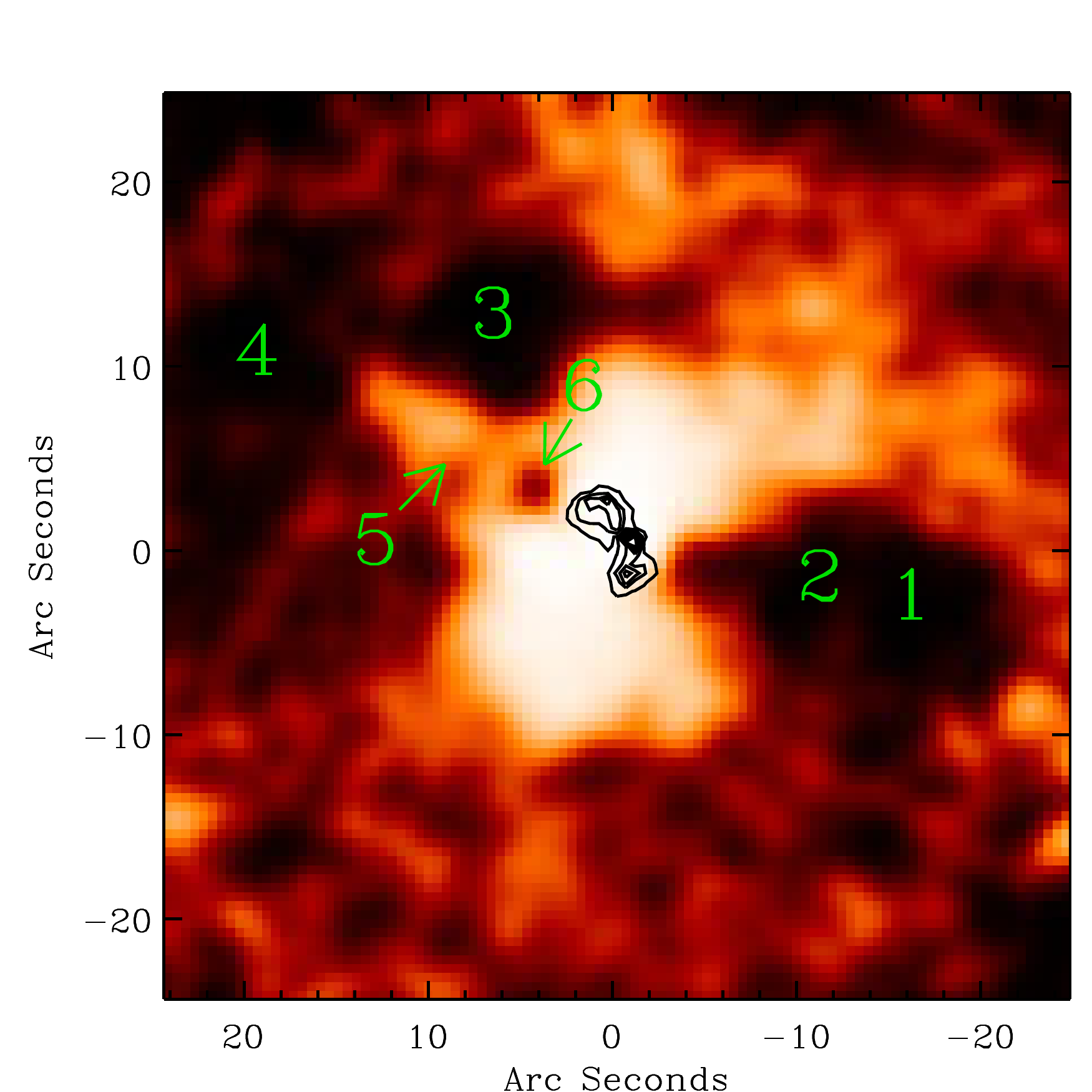}
\end{center}
\caption{ Two unsharp mask images of the merged 150 ksec {\it Chandra}
  0.5-7  keV  observation.  The  panel  at {\it  left}  was  made  via
  subtraction of  a 20\arcsec Gaussian  smoothed version of  the image
  from the adaptively smoothed version. The version in the {\it right}
  panel was made by  subtracting the same 20\arcsec\ Gaussian smoothed
  version  of the image  from a  5\arcsec\ Gaussian  smoothed version,
  then dividing  by the sum  of the two  images.  In the  chosen color
  scheme,  regions  of  X-ray   surface  brightness  excess  over  the
  subtracted  smoothed  counts appear  in  red/orange, while  deficits
  appear in  black.  330 MHz VLA  contours from have  been overlaid in
  green, while the 1.3 GHz radio contours are plotted in blue, and the
  8.4 GHz  radio contours appear  in black. All  of the radio  data is
  from  \citet{clarke05}.  Note  how  the extended  330  MHz  emission
  corresponds  in  both linear  extent  and  projected  P.A. with  the
  apparent X-ray cavity consisting of features (1) and (2), as labeled
  in the  right panel.  Note also  that the northeast hook  of the 1.3
  GHz emission  (blue) is  aligned with the  ``bottom'' of the  15 kpc
  X-ray filament.  The major  morphological features which will be the
  subject of further spatial and spectral analysis have been labeled 1-6 in the  right panel. Although we
  label them separately, we do not mean to imply that features 1 and 2
  are necessarily discrete. Rather, feature 1 is labeled separately to
  enable  a  clearer comparison  with  the \citet{mcnamara01}  ``ghost
  cavity''. Features 1  and 2 may well be one  larger cavity. The ring
  features seen in the  left-hand panel are artifacts from subtracting
  images with different smoothing lengths. }
\label{fig:unsharp}
\end{figure*}

\section{X-ray Spatial and Spectral Results}
\label{section:xrayspatial}

\subsection{General X-ray morphology}
\label{section:generalxraymorphology}

Previous X-ray studies have been published for A2597 using the 40 ksec
(18 ksec  of good time) {\it  Chandra} observation 922  (e.g., M01 and
\citealt{clarke05}, hereafter  C05).  M01 was  the first to  report on
the western  and northeastern  ghost cavities inferred  from localized
depressions  in  the  X-ray  surface  brightness. Later  work  by  C05
extended this analysis to include a study between X-ray morphology and
Very Large Array (VLA) radio observations at 8.4 GHz, 1.3 GHz, and 330
MHz.   That  work described  an  X-ray ``tunnel''  apparent in  their
unsharp  mask  of  the  922  data.  Roughly  22\arcsec\  (33  kpc)  in
projected  length and  1\farcs5  (2.3 kpc)  in  projected radius,  the
tunnel appeared  to connect the  peak of the X-ray  surface brightness
distribution, cospatial with the radio  and BCG optical core, with the
M01 western cavity.   It was not apparent whether  this was a discrete
feature, or was  part of a larger cavity  encompassing both the tunnel
and the M01  cavity.  Extended 330 MHz radio  emission was observed to
be cospatial with both features.

\begin{figure*}
\begin{center}
\includegraphics[scale=0.47]{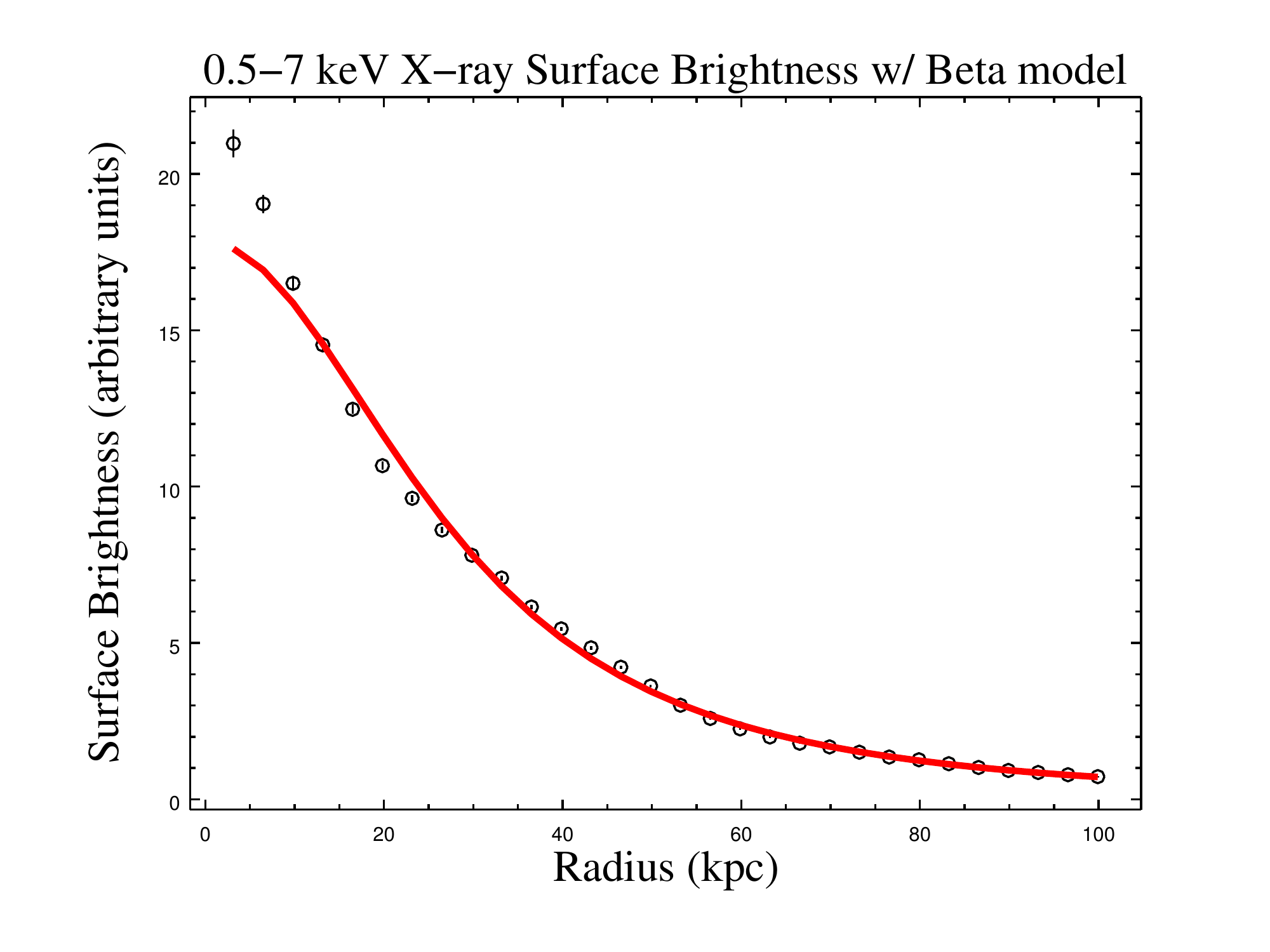}\hspace*{-6mm}\includegraphics[scale=0.47]{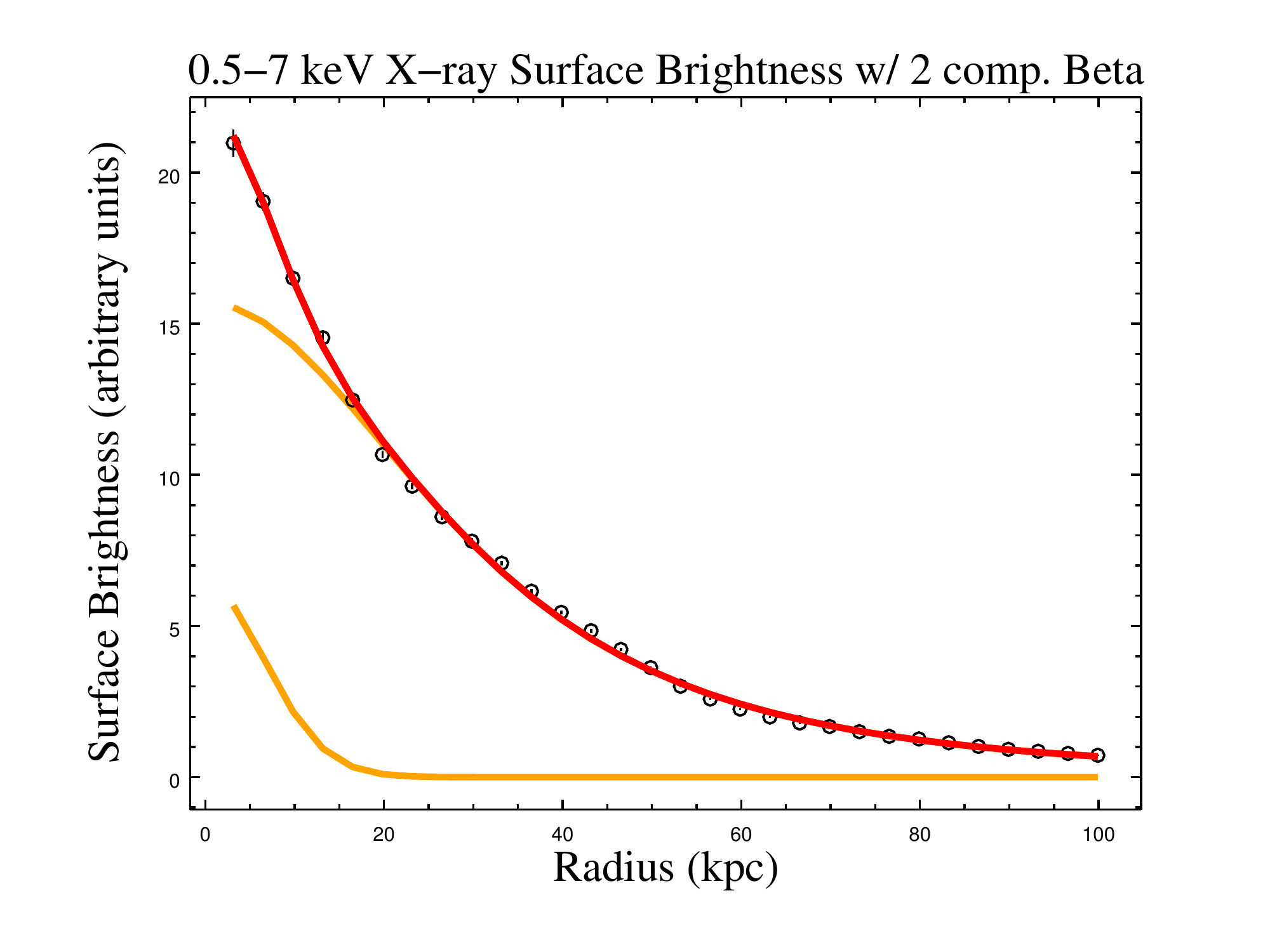}
\end{center}
\caption{X-ray surface brightness profiles extracted from a series of concentric annuli
spanning from the central regions out to 100 kpc. ({\it left}) A one
component beta model has been fit.  ({\it right}) A two-component (additive)
beta model has been fit, yielding better overall correspondence to the measured profile. The additive two-component model is in red, while the two individual components to the model are shown in orange. 
}
\label{fig:surfacebrightness}
\end{figure*}

In Fig.~\ref{fig:chandra} we present the new, combined 0.5-7 keV data.
The left  panel shows  the exposure corrected  mosaic with  no spatial
binning or  smoothing applied. The  right panel shows the  merged data
adaptively smoothed  with a variable width gaussian  kernel whose size
self-adjusts  to   match  the  local  event   density.  Black  surface
brightness contours are overlaid to make individual features easier to
view. The innermost contour marks a flux of $4.6\times10^{-7}$ photons
sec\mone\ cm\mtwo\  pixel\mone, and move  outward with a  decrement of
$2.0\times10^{-8}$ photons sec\mone\  cm\mtwo\ pixel\mone. Both panels
share   the   same   scale,   with   a  field   of   view   (FOV)   of
$\sim60\arcsec\times60\arcsec$,  corresponding to  90 kpc  $\times$ 90
kpc. The FOV  is centered on the X-ray centroid  at RA=23h 25m 19.75s,
Dec =-12$^\circ$ 07' 26.9'' (J2000).  As with all figures presented in
this paper, North is up, and East is left.

The  X-ray  surface
brightness  distribution  is  highly anisotropic,  and extended along a position angle that matches the major
axis  of  the  BCG  stellar   isophotes  (which  lie  on  a  P.A.$\sim
-45^\circ$, where North through East, or counter-clockwise, is the
positive  direction).  The  spatial anisotropy  of the  X-ray emission
seems to be  confined to the scale of the BCG,  as the X-ray isophotes
assume  a smoother  elliptical  shape in  the  outermost regions.

As  seen in  Fig.~\ref{fig:chandra}, the  innermost 20\arcsec\  of the
0.5-7 keV emission is distributed  in a butterfly-like shape, with two
high surface  brightness knots $\sim1\arcsec$ W  and $\sim3\arcsec$ SE
of the X-ray centroid. The knot 1\arcsec\ W is approximately cospatial
with the  radio core. A  fraction of the  X-ray emission in  this knot
stems from the  weak point source associated with  the AGN.  Two sharp
deficits  of X-ray  emission $\sim2\arcsec$  NE and  $\sim2\arcsec$ SW
define the  inner ridges of the  ``butterfly wings''. One  of the most
prominent  features visible in  Fig.~\ref{fig:chandra} is  a 10\arcsec
($\sim   15$   kpc  in   projection)   filament   extending  along   a
P.A.$\approx55^\circ$, roughly perpendicular to  the major axis of the
butterfly feature and aligned (in  projection) along the minor axis of
the  BCG stellar isophotes.  The western  edge of  the M01
ghost cavity is  faintly seen $\sim 18\arcsec$ W,  $5\arcsec$ S of the
centre.

In Fig.~\ref{fig:unsharp} we show the same 0.5-7 keV data processed in
two   ways:   (1)   a   residual   image   made   by   subtracting   a
30\arcsec\ gaussian smoothed image  from the adaptively smoothed image
shown in  Fig.~\ref{fig:chandra}{\it b},  and (2) a  more conventional
unsharp mask made by  subtracting a 20\arcsec\ gaussian smoothed image
from  a  5\arcsec\  (non-adaptive)  gaussian  smoothed  version.   The
subtracted data is then divided by the sum of the two images.  Regions
of  X-ray  surface  brightness   excess  over  the  subtracted  smooth
background  appear  in  white,  while deficits  (cavities)  appear  in
black. Both  panels are  on the same  scale, with  identical 50\arcsec
$\times$  50\arcsec\ (75  kpc$\times$75  kpc) FOVs.  We overlay  radio
contours on  the left panel, which  we will discuss  later.  These two
edge enhancement methods (particularly method 1) are inherently noisy,
and  introduce artifacts  that  complicate quantitative  morphological
analysis.  Significance of  the  observed features  must therefore  be
estimated from the un-processed data.  We show these mostly as viewing
aids for the major morphological features.

By  comparing unsmoothed  counts in  like  sized regions  at the  same
cluster-centric radius, we find  six discrete features associated with
$\gae10\sigma$ deficits or excesses relative to the local mean.  These
features    are    labeled    1-6    in    the    right    panel    of
Fig.~\ref{fig:unsharp}. This  paper discusses feature  (5), the ``cold
filament'',  while  energetics  analysis  of  the  X-ray  cavities  is
discussed in a companion paper on AGN heating and ICM cooling in A2597 \citep{tremblay_cooling}.

\begin{figure*}
\begin{center}
\includegraphics[scale=0.26]{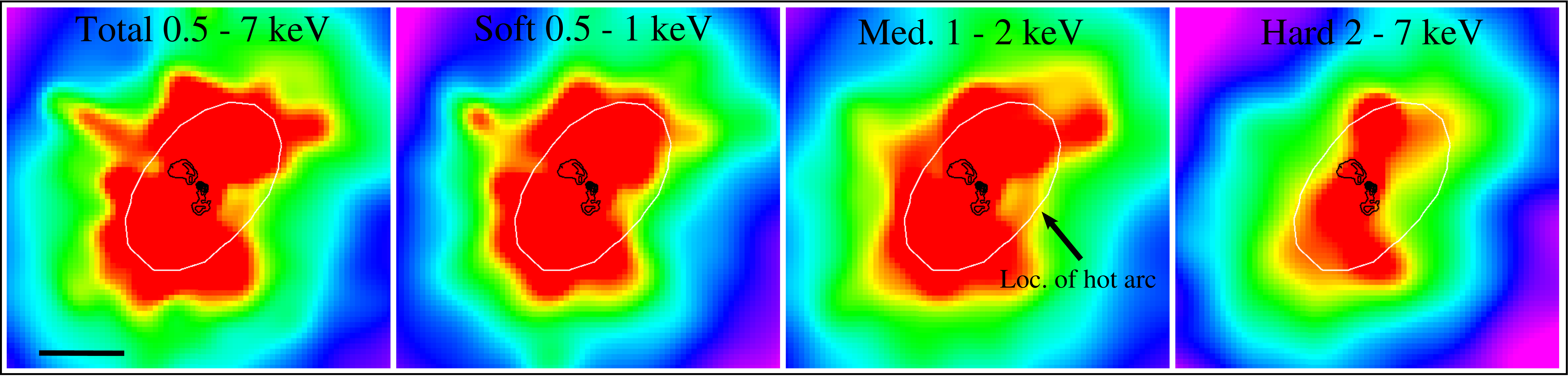}
\end{center}
\vspace*{-3mm}
\caption{The adaptively smoothed {\it Chandra} image of A2597 shown at
  various  slices in  energy  space, including  (from  left to  right)
  ``total'' (0.5-7  keV), ``soft'' (0.5-1 keV),  ``medium'' (1-2 keV),
  and ``hard'' (2-10 keV) emission.  In black contours, we overlay the
  8.4 GHz  radio contours  of PKS 2322-122,  and in white  contours we
  overlay a stellar isocontour from the {\it HST} $R$-band observation
  of the host galaxy. The black  line at the bottom left corner of the
  leftmost panel marks a distance of 10\arcsec (15 kpc). Note that the
  major axis of the host  galaxy stellar isophote is aligned along the
  same position  angle as  the X-ray major  axis. The  bright filament
  which extends 15 kpc NE from the center in the total and soft panels
  is effectively absent in the  rightmost hard panel.  The filament is
  therefore a soft excess, colder  than the surrounding gas.  The hard
  X-ray  ``disk''  evident  in  the  rightmost  panel  is  aligned  in
  projection  with  the isophotal  major  axis  of  its host  galaxy's
  stellar component.  }
\label{fig:hardness}
\end{figure*}

\subsection{Radio \& X-ray spatial correlations}

In Fig.~\ref{fig:unsharp}{\it a} we overlay  8.4 GHz, 1.3 GHz, and 330
MHz VLA radio  contours in black, blue, and  green (respectively).  We
do not  show the innermost 1.3 GHz  and 330 MHz contours  to allow the
higher frequency contours to be  clearly visible.  As noted by M01 and
C05,  the northern and  southern lobes  of the  8.4 GHz  source (black
contours)  are bounded  to  the west  and  east by  deficits of  X-ray
emission associated  with features (6) and  (2), respectively.  Beyond
this, it is difficult to discern  whether or not the 8.4 GHz lobes are
interacting with  the keV gas on  these scales. No  X-ray cavities are
observed to be directly cospatial  with the 8.4 GHz lobes, though this
lack of evidence should not be interpreted as evidence of absence. Any
cavities that  may exist  on these scales  would be very  difficult to
detect given a combination of  (a) the compactness of the radio source
compared to the X-ray emission and (b) a great deal of intervening keV
gas along  the line of sight.  We expect that  any cavities associated
with the  8.4 GHz  lobes would be  associated with  surface brightness
decrements  of  a  few   percent  at  most,  strongly  limiting  their
detectability.    Past    literature    presents    strong    evidence
\citep[e.g.,][]{koekemoer99,odea04}  that   the  8.4  GHz   source  is
dynamically interacting  with the $\sim 10^4$ K  emission line nebula,
which we will discuss later.

The 330  MHz VLA A-configuration observation (in  green contours) does
not resolve  structures at the scale  of the 8.4 GHz  source, though a
prominent arm of emission extends $\sim70$ kpc to the west (projected)
in strong  spatial correlation with the western  large cavity (feature
1+2). It  is worth noting that the  symmetric 50 pc scale  jets in the
VLBA  map  of   \citet{taylor99}  are  approximately  aligned  (within
$\sim5^\circ$) with the major axis of the western large cavity and the
extended 330 MHz emission (though this of course involves a comparison
at vastly different  scales).  The blue 1.3 GHz  contours are extended
along the  same axis as the  330 MHz eastern  extension, the projected
VLBA jet axis, the western large cavity, as well as $\sim 8$ kpc along
the bottom half of the 15  kpc X-ray filament (feature 5).  The NE
1.3 GHz hook, which is $\sim  15$ kpc from the radio core, also covers
the C05 filament base cavity,  (feature 6).  There is also a $\sim
1$ kpc extension to the SW, into the bottom opening of the western
large cavity.

The projected spatial  coincidence of all of these  features is strong
evidence for a common, current jet axis. While the P.A. of the 8.4 GHz
source is  offset from this axis,  it is the exception,  and may arise
from dynamical frustration of the  jet as it propagates amid the dense
molecular  medium harboring the  emission line  nebula (this  has been
discussed   by  e.g.,  \citealt{sarazin95,koekemoer99,odea04,oonk10}).
The 1.3  GHz emission  extended along  the bottom half  of the  15 kpc
X-ray filament  (feature 5 in Fig.~\ref{fig:unsharp})  may be evidence
for  dredge-up of  lower  temperature,  high density  keV  gas by  the
propagating  radio  jet. \citet{oonk10}  reported  high velocity  (and
velocity dispersion) streams of  H$_2$ and \ion{H}{ii} coincident with
the southern edge of the northern 8.4 GHz radio lobe and approximately
aligned with  the projected VLBA jet  axis.  The jet  may therefore be
dynamically interacting with both the  hot and warm/cold phases of the
ISM.  We will investigate these possibilities in Section 5.

\subsection{X-ray Surface brightness Profile}

In Fig.~\ref{fig:surfacebrightness} we plot azimuthally summed X-ray surface brightness 
against projected radius, using fluxes extracted from narrow concentric annuli
spanning $\sim5-100$ kpc. We fit
both singular and  additive two-component Lorentz
1-D ``beta'' models with a varying power law (\texttt{beta1d} in CIAO Sherpa) to the
data, wherein the surface brightness $\Sigma_\mathrm{X}$ at projected radius
$r$ varies as 
\begin{equation} 
\Sigma_\mathrm{X}\left(r\right) =
\Sigma_{\mathrm{X}}\left(0\right) \left[ 1 + \left(\frac{r}{r_0}\right)^2\right]^{\left(-3
\beta + 1/2\right)}. 
\end{equation} 
Here, $\Sigma_{\mathrm{X}}\left(0\right)$ is the central
X-ray surface brightness and $r_0$ is the core radius. 
Typically, the ICM in galaxy clusters is well-fit by models with $\beta\approx0.6$ \citep{sarazin86}. 
Our results are consistent with this: if we
fit a one-component beta model, shown in the left panel of Fig.~\ref{fig:surfacebrightness}, we find $\beta=0.633$ with core radius
$r_0=16\farcs4$ (or $\sim24$ kpc). A far better fit is obtained with an additive
two-component beta model, which we plot in the right panel in Fig.~\ref{fig:surfacebrightness}
(red line). This model consists of a $\beta=0.686$ fit with a core radius of 
$r_0=19\farcs1$ (or $\sim29$ kpc) and an inner steep component ($\beta=10$) to fit the cusp (yellow lines). 
The small deviations between the model and data arise (in part) because the annuli from which the spectra are 
extracted effectively give an azimuthal average
of X-ray fluxes that vary strongly as a function of position angle, owing to the morphologically 
disturbed nature of the X-ray gas. In the plots, errors bars are shown, though 
difficult to see because they are generally smaller than the data points.

\begin{figure*}
\begin{center}
\includegraphics[scale=0.66]{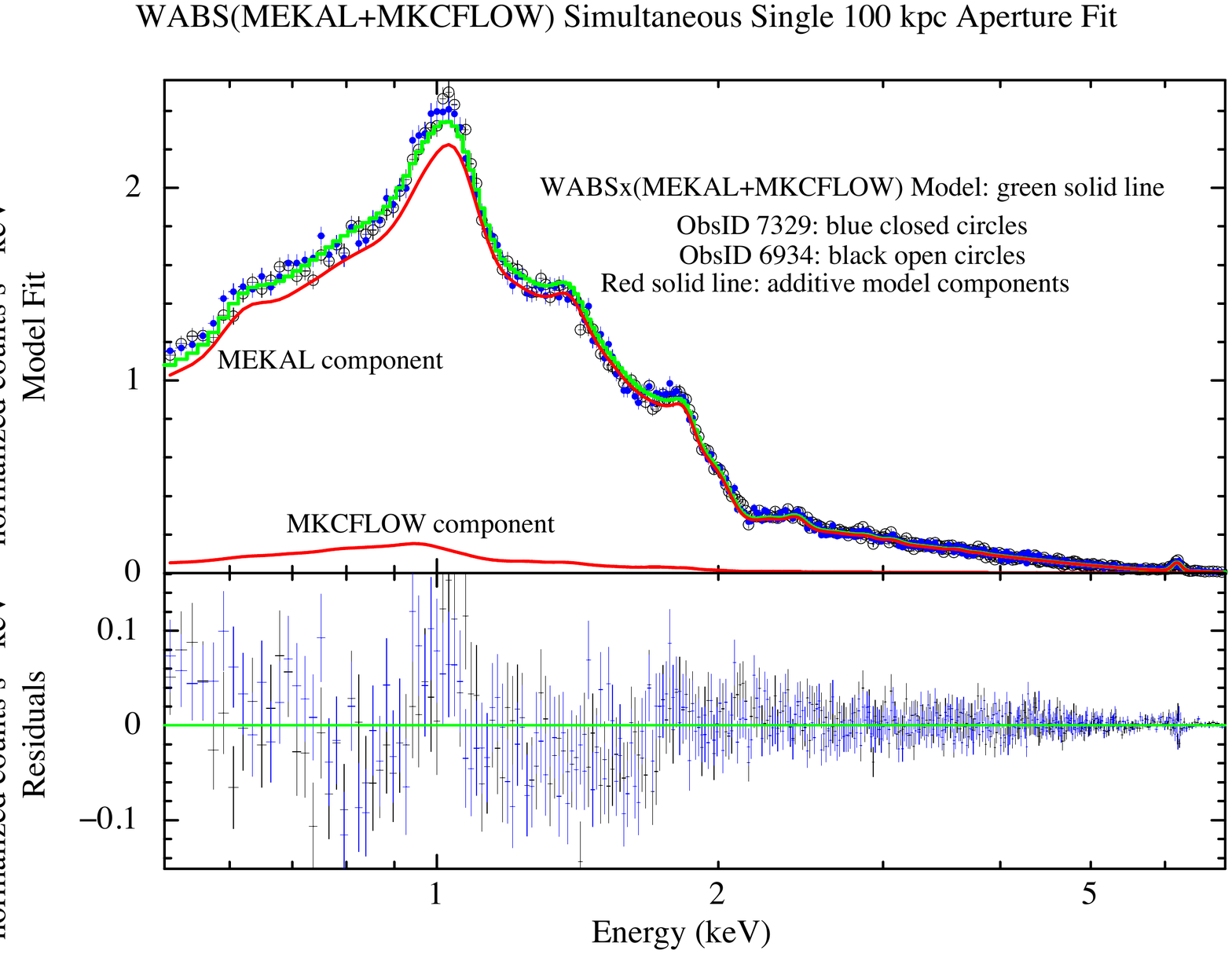}
\vspace*{-8mm}
\end{center}
\caption{0.5-7 keV X-ray spectrum of the innermost 100 kpc of the Abell 2597 BCG, extracted from {\it Chandra} observations 6394 (black open
circles) and 7329 (blue closed circles). The spectra have extracted independently, grouped using a 30 count bin size, background subtracted, and fit simultaneously after removal of contaminating sources. 
A reduced cooling flow model (green solid line) has been fit to the data (\texttt{WABS} $\times$ [ \texttt{MEKAL} + \texttt{MKCFLOW}] in \texttt{XSPEC}). The individual additive model components (\texttt{MEKAL} and \texttt{MKCFLOW}) are shown in red solid lines and labeled accordingly. The high temperature parameter in the \texttt{MKCFLOW} component has been tied to the \texttt{MEKAL} ambient temperature component to 
simulate a residual cooling flow proceeding at low levels from the hot X-ray atmosphere.   
Fit residuals are shown in the 
bottom panel using the same color coding.  They reach an $\sim 8$\% maximum at the Fe L 1 keV peak. Both panels are in units of normalized counts s\mone\ keV\mone. 
See the last entry in Table \ref{tab:totalxrayfits} for model parameters associated with this fit. Note that other model fits presented in the same table yield similar results in terms of reduced chi squared, illustrative of the degeneracy inherent in this sort of X-ray spectral fitting. }
\label{fig:mekal}
\end{figure*}

\begin{table*}
\centering
\caption{Simultaneous spectral fits to the 0.5-7 keV X-ray spectrum extracted from a central 100 kpc aperture. Prior to extraction, a temporal filter 
was applied to the data to reject flaring events, and compact sources associated with other cluster members as well as the central weak 
point source associated with the AGN was removed. The spectra were rebinned using a 30 count threshold. Observation 922 has been excluded from spectral analysis 
due to significant flaring events and high background levels.  Symmetric one sigma confidence intervals are shown on fit parameters. Parameters that have been frozen to a 
specific 
value prior to fitting are shown in parentheses. In the last model, \texttt{WABS} $\times$ [ \texttt{MEKAL} + \texttt{MKCFLOW}], 
we tie the high temperature \texttt{MKCFLOW} parameter to the \texttt{MEKAL} $kT$ parameter, following similar ``reduced'' cooling flow modeling 
in the literature (e.g., \citealt{rafferty06,russell10}).  }
\begin{tabular}{cccccccc}
\hline
 &
$N_\mathrm{H}$ &
$kT_\mathrm{low}$ &
$kT_\mathrm{high}$ &
Abundance &
\texttt{MEKAL} Norm. &
\texttt{MKCFLOW} $\dot{M}$ &
red.~chi sq. \\
XSPEC Model &
($\times10^{20}$ cm$^{-2}$) &
(keV) &
(keV) &
(solar) &
($\times 10^{-2}$) &
$(M_\odot$ yr$^{-1})$ &
$\left(\chi^2/\mathrm{dof}\right)$ \\
(1) & (2) & (3) & (4) & (5) & (6) & (7) & (8) \\
\hline
\hline
\texttt{WABS} $\times$ \texttt{MEKAL} & $1.5\pm0.12$    &  \nodata   &  $3.42\pm0.02$   & $0.43\pm0.01$  &  $1.501\pm0.009$ & \nodata & $745/735=1.014$\\
\texttt{WABS} $\times$ \texttt{MEKAL} & $1.63\pm0.19$   &  \nodata & $3.39\pm0.04$   &   (0.4)  &  $1.524\pm0.008$  &  \nodata &  $754/736=1.025$ \\
\texttt{WABS} $\times$ \texttt{MEKAL} & (2.48)   &  \nodata & $3.31\pm0.03$   &   (0.4)  &  $1.55\pm0.005$  &  \nodata &  $807/737=1.096$ \\
\texttt{WABS} $\times$ [ \texttt{MEKAL} + \texttt{MEKAL}]   &  (2.48)  &  $2.68\pm0.13$  &  $6.01\pm1.05$ & (0.4) &   $1.10$\&$4.57\pm2.06$ & \nodata & $759/735=1.032$ \\
\texttt{WABS} $\times$ \texttt{MKCFLOW} & (2.48)   &  $1.88\pm0.07$  &  $5.33\pm0.16$   & (0.4)   & \nodata   &  $497\pm33$  & $763/736=1.04$ \\
\texttt{WABS} $\times$ \texttt{MKCFLOW} &  $1.59\pm0.17$    &    $1.87\pm0.08$   &   $5.61\pm0.17$   &   (0.4) & \nodata  & $454\pm29$  & $707/735=0.96$ \\
\texttt{WABS} $\times$ [\texttt{MEKAL} + \texttt{MKCFLOW}]  & (2.48)  & (0.1)  &  $3.47\pm0.05$ & $0.43\pm0.018$ & $1.47\pm0.0002$  & $22 \pm 5$ & $740/739=1.008$ \cr
\hline
\end{tabular}
\label{tab:totalxrayfits}
\end{table*}

\begin{figure*}
\includegraphics[scale=0.5]{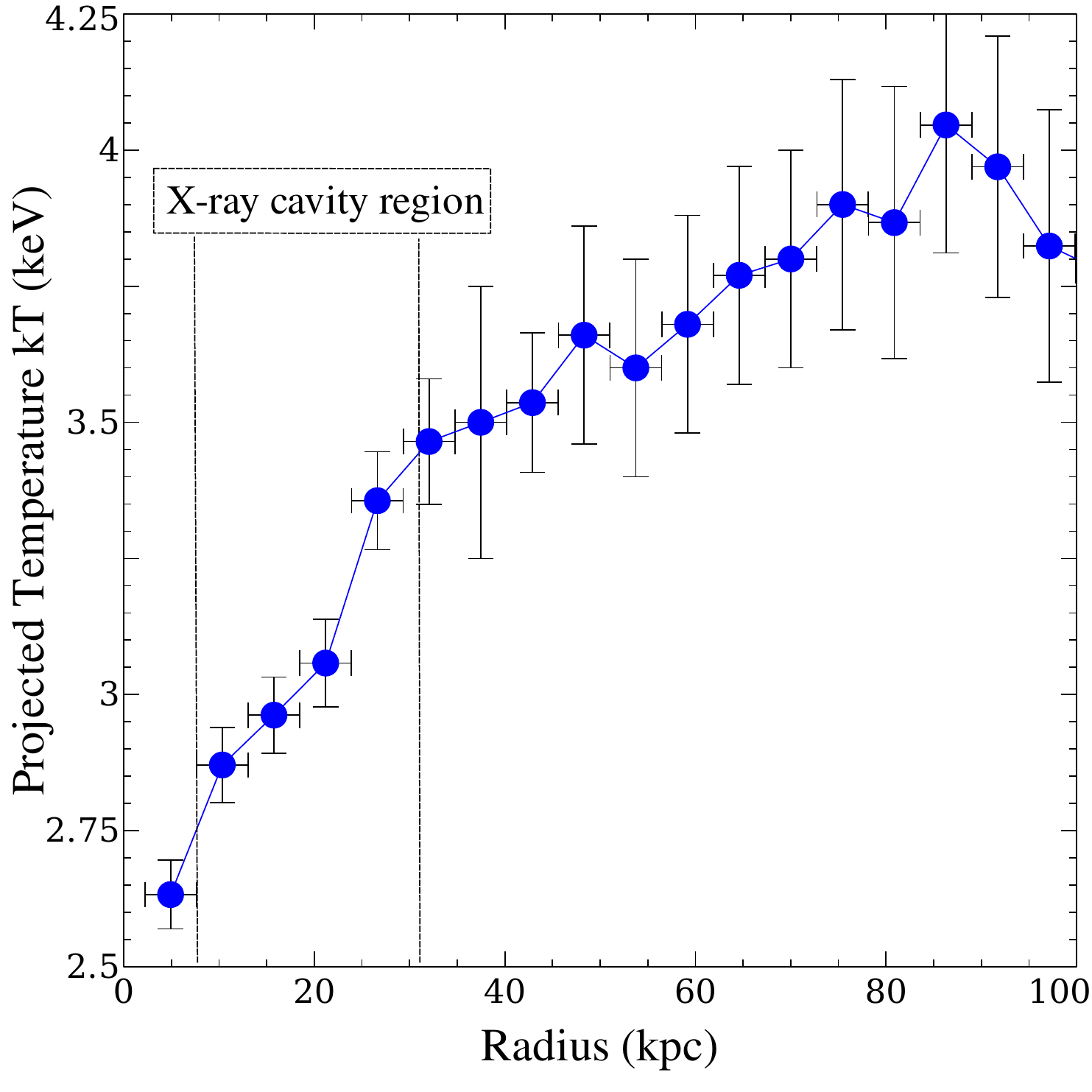}\includegraphics[scale=0.5]{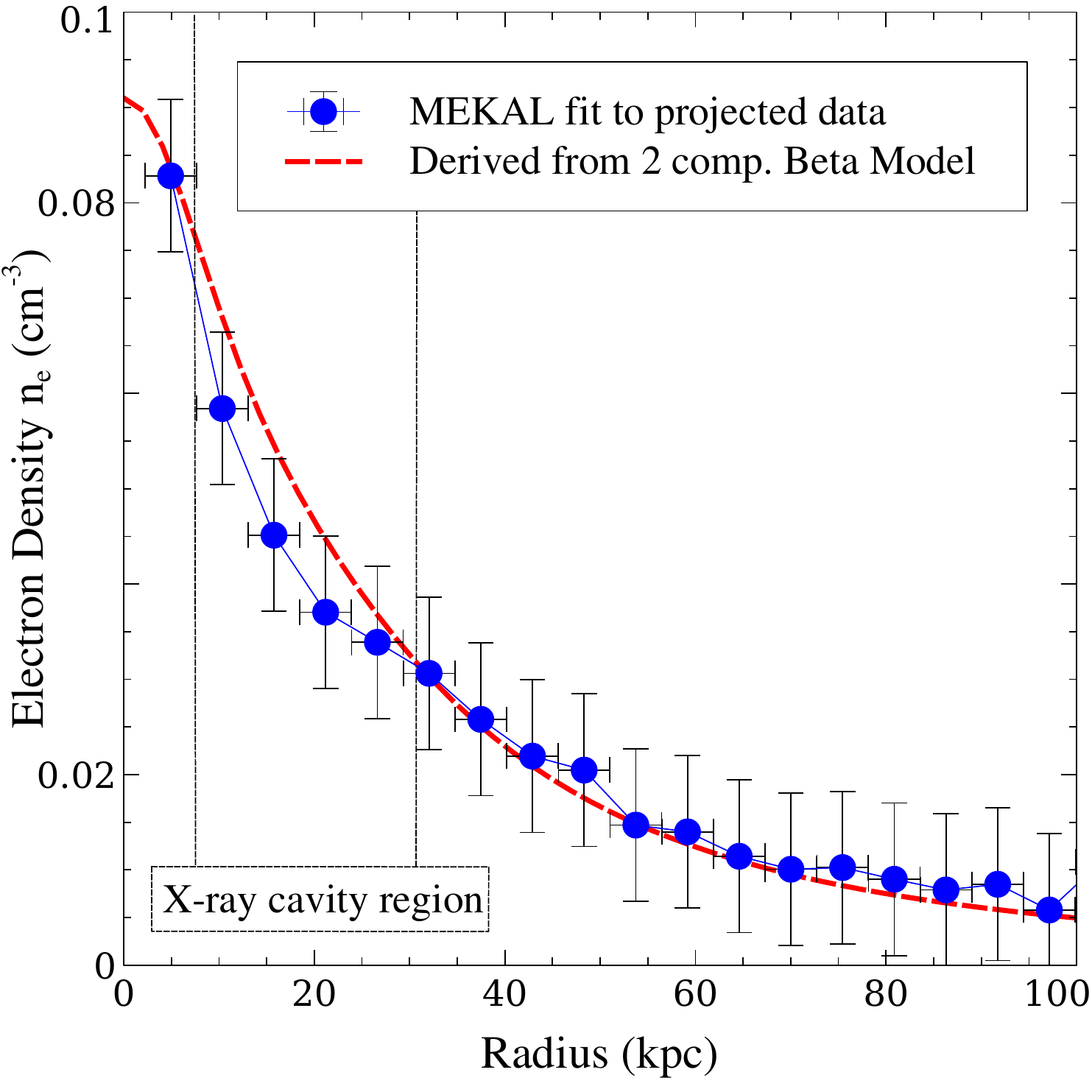}
\includegraphics[scale=0.53]{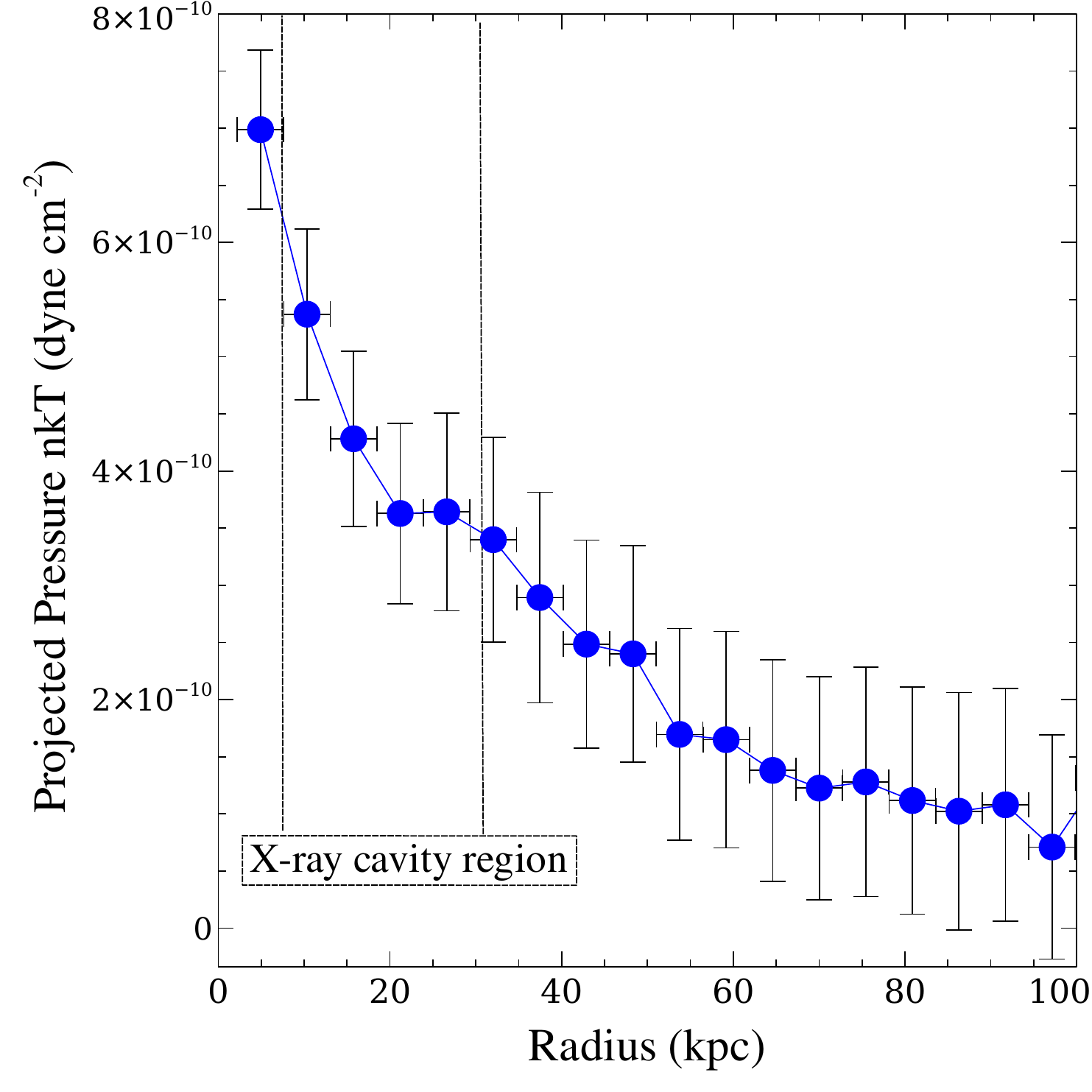}\hspace*{-4mm}\includegraphics[scale=0.53]{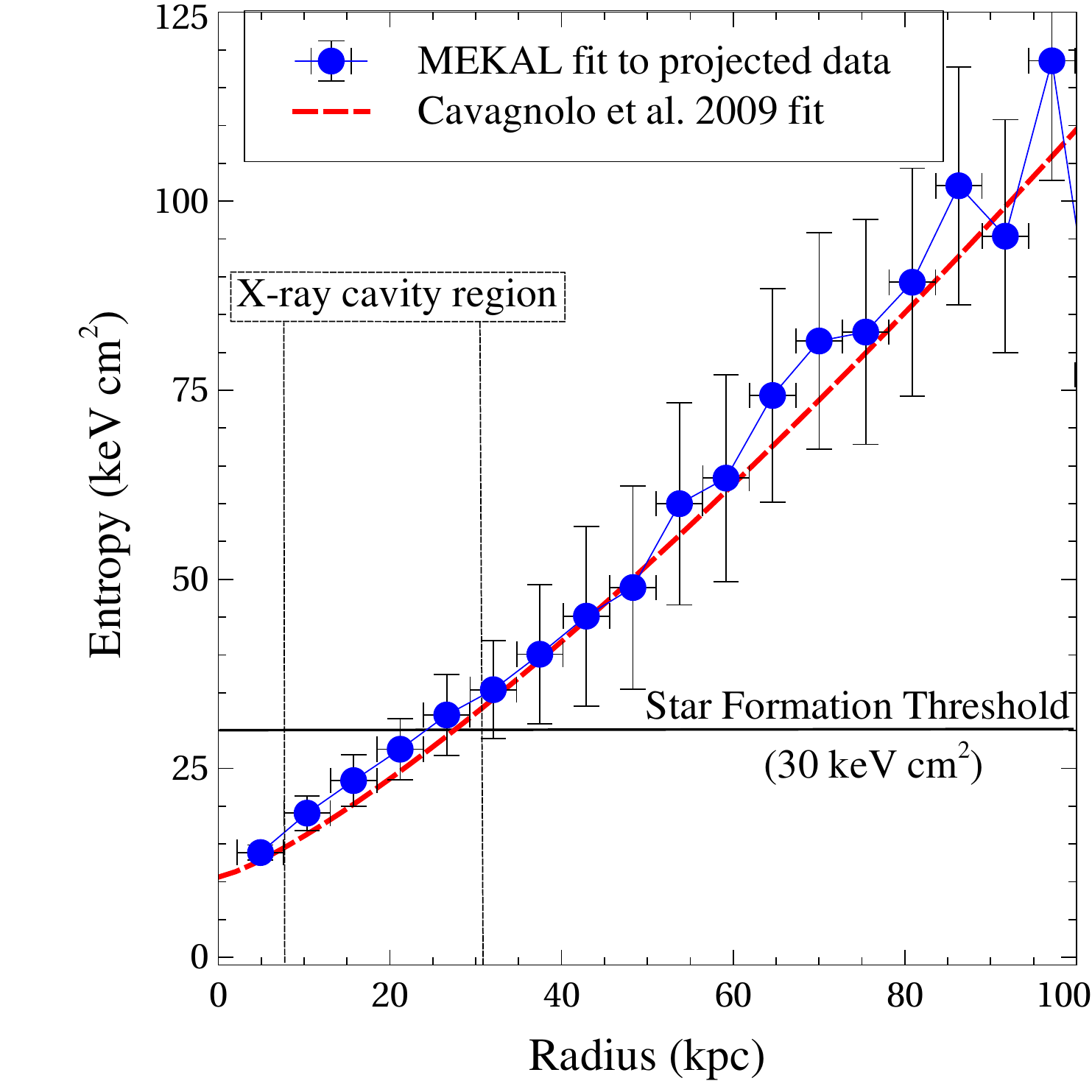}
\caption{({\it top left}) Projected temperature and ({\it top right}) projected density profiles derived 
from \texttt{MEKAL} thermal plasma models fit to radially extracted spectra from the inner 
100 kpc of A2597.  
The temperature $kT$ is a fit parameter from the \texttt{MEKAL} model. The electron 
density ($n_e$) was derived from the \texttt{MEKAL} normalization, while the pressure ($P$) and entropy ($S$) profiles, 
shown at  ({\it bottom left}) and  ({\it bottom right}), 
were calculated using  $P=nkT$ (assuming $n=2n_e$) and  $S=kTn_e^{-2/3}$, respectively. 
In each profile we mark the $\sim 25$ kpc region permeated by the X-ray cavity network. 
Independent consistency checks on the electron density and entropy profiles are plotted in red. 
In the electron density plot, the consistency check is from the two-component additive beta model 
fit to the X-ray surface brightness profile. In the entropy profile, we plot the best-fit entropy profile 
from \citet{cavagnolo09}. }
\label{fig:deprojection}
\end{figure*}

\subsection{0.5-7 keV spectrum from a central 100 kpc aperture}

We extract a spectrum from the individual flare-filtered ObsIDs 6934 and 7329, 
using a  single circular aperture with a 63\farcs3 (100 kpc) radius encompassing most of
the sharply peaked central X-ray emission. 
Prior to this, 
we removed contaminating point sources.  The spectra were extracted using the {\sc ciao} 4.2 script
\texttt{specextract}, and count-weighted response matrices were generated for
each extraction.

Using simultaneous fitting techniques in XSPEC,
we fit a variety of models to the total 0.5-7 keV spectrum extracted from the
100 kpc aperture. 
The most simple of these is a \texttt{MEKAL} model that has been absorbed to account for
attenuation by the galactic hydrogen column (\texttt{WABS} $\times$
\texttt{MEKAL}).  We also fit absorbed two-component \texttt{MEKAL} models
(\texttt{WABS} $\times$ [ \texttt{MEKAL} + \texttt{MEKAL}]) as well as the
standard cooling flow model in XSPEC (\texttt{WABS} $\times$ \texttt{MKCFLOW}).
Finally, we fit a ``reduced'' cooling flow model \texttt{WABS} $\times$ [ \texttt{MEKAL} + \texttt{MKCFLOW}]  (e.g., \citealt{rafferty06}), 
in which the high temperature \texttt{MKCFLOW} parameter is tied to the \texttt{MEKAL} $kT$ parameter.
The physical motivation behind this model is that AGN heating maintains the bulk of the gas 
at the ambient temperature (fit by the \texttt{MEKAL} component), while the \texttt{MKCFLOW} component models residual cooling 
from that ambient temperature (hence the parameter sharing) below 0.1 keV (to which $kT_\mathrm{low}$ is fixed).  
In all cases, the source redshift was fixed to $z=0.0821$.

Results from our modeling are shown in 
Table~\ref{tab:totalxrayfits}.  As evident in column (8) of that table, the
fits are uniformly ``good'' in terms of reduced $\chi^2$, indicative of the
degeneracy that almost always arises from this type of X-ray spectral fitting.
The last line of this table lists results from fitting the ``reduced'' cooling flow model, \texttt{WABS} $\times$ [ \texttt{MEKAL} + \texttt{MKCFLOW}], which has been used frequently in recent literature \citep[e.g.,][]{russell10}. We show this fit in Fig.~\ref{fig:mekal} (green line). 
The individual additive model components (\texttt{MEKAL} and \texttt{MKCFLOW}) are shown in red solid lines and labeled accordingly. The high temperature parameter in the \texttt{MKCFLOW} component has been tied to the \texttt{MEKAL} ambient temperature component to 
simulate a residual cooling flow proceeding at low levels from the hot X-ray atmosphere.   
Fit residuals are shown in the 
bottom panel using the same color coding.  They reach an $\sim 8$\% maximum at the Fe L 1 keV peak.  
We discuss the residual cooling flow model for A2597 at greater length in \citet{tremblay_cooling}.

\begin{figure*}
\begin{center}
\includegraphics[scale=0.5]{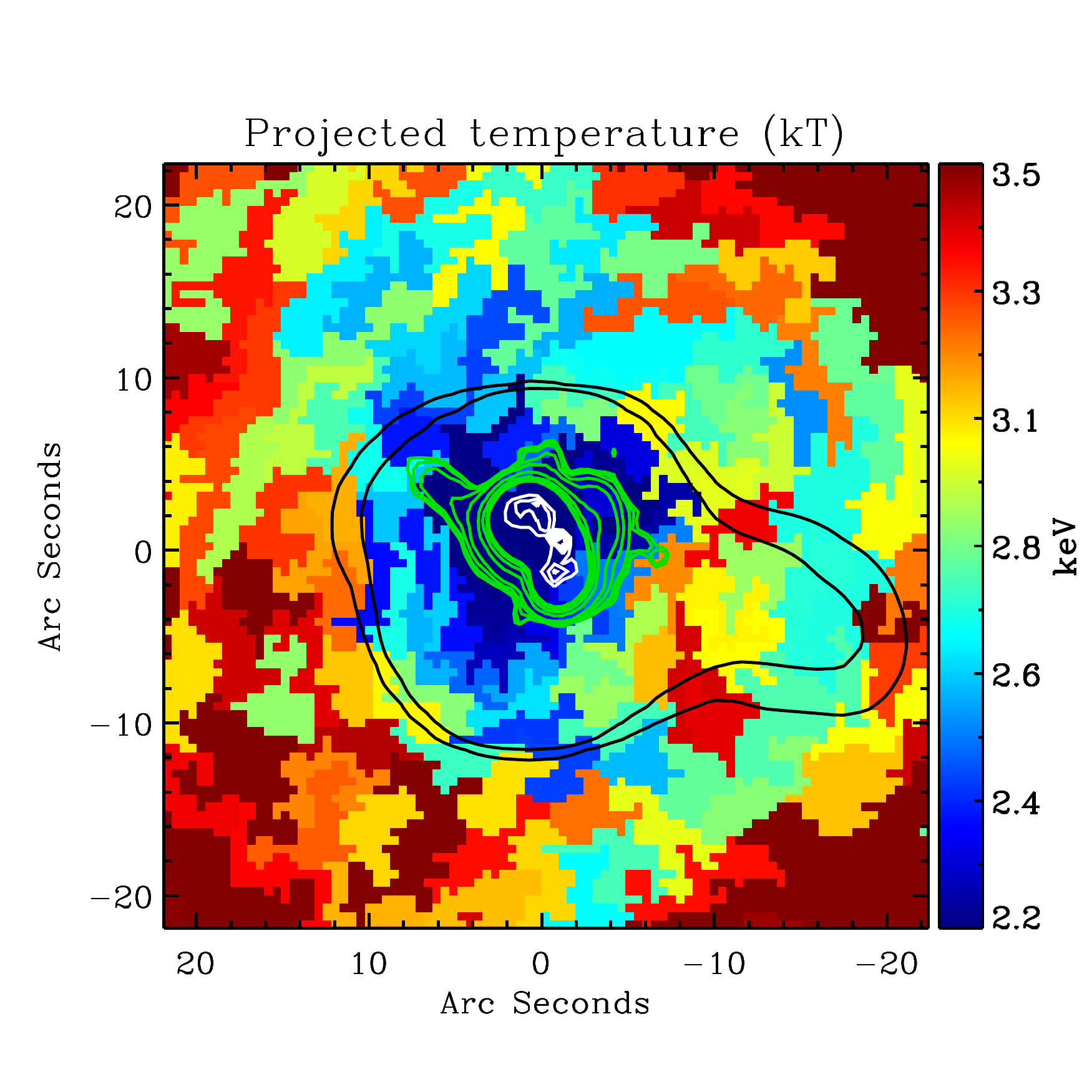}\hspace*{-3mm}
\includegraphics[scale=0.5]{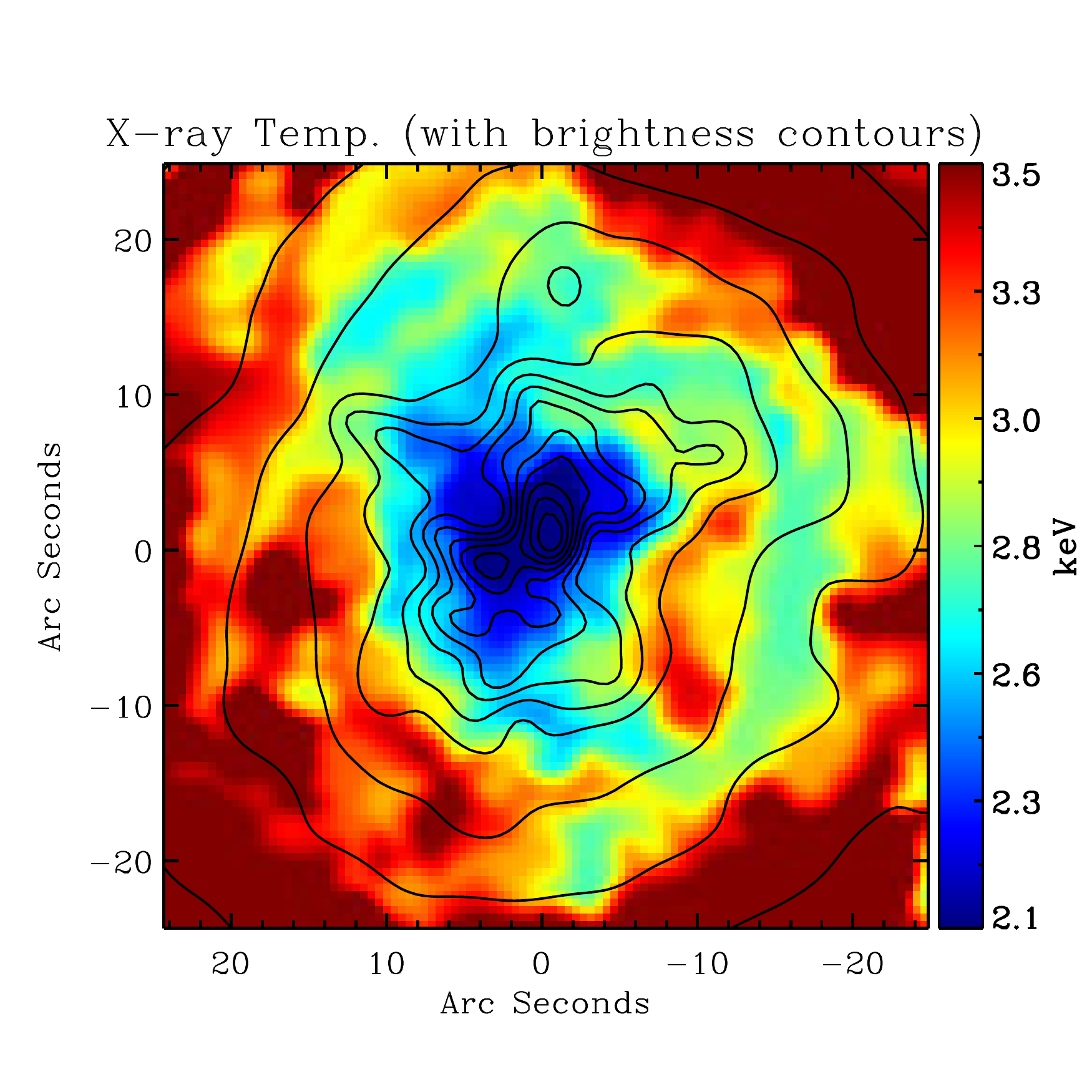}
\end{center}
\vspace*{-10mm}
\caption{({\it  left}) X-ray  temperature map  created  from spatially
  resolved {\it  Chandra} spectroscopy  of A2597. The  best-fit single
  \texttt{MEKAL} model temperature is  encoded via the scaled colorbar
  at the  left of  the figure. Note  the 10\arcsec\  ``cold filament''
  extending from the central cold  ($\sim2.2$ keV) feature, as well as
  the  ``hot''  arc-shaped  feature  10\arcsec\  to  the  west.  These
  features and their associated physical interpretations are discussed
  in the text.  ({\it right}) Adaptively smoothed version  of the same
  X-ray  temperature  map.   X-ray  surface  brightness  contours  are
  overlaid in black.  At  the redshift of A2597, 1\arcsec\ corresponds
  to $\sim1.5$ kpc. East is left, north is up.}
\label{fig:full_tmap}
\end{figure*}

\subsection{Hardness analysis}

\label{section:hardness}

In Fig.~\ref{fig:hardness} we show  soft (0.5-1 keV) through hard (2-7
keV) cuts  in energy space  for the adaptively smoothed  combined {\it
  Chandra} data.  In each panel  we overlay the  8.4 GHz radio  map in
black contours,  which for reference is $\sim10$  kpc from end-to-end.
The  previously  mentioned  15  kpc  NE filament  becomes  fainter  at
successively harder slices in energy space, and effectively disappears
in  the rightmost  panel. The  filament is  therefore a  relative soft
excess,  and is likely  to be  colder than  the surrounding  gas.  The
butterfly feature  also loses  much of its  NE-SW width  at higher
energies  while retaining  effectively  the same  extension along  the
NW-SE axis  that is  aligned with  the major axis  of the  BCG stellar
isophotes. We have also created a hardness ratio map, but this will be
discussed in the following section.
In the 1-2 keV panel, we have marked the location of the ``hot arc'', a feature
which will be discussed in Sections 4 and 6.

We note that soft X-ray emission along lines of sight
that pass  through the galaxy  midplane suffers more  attenuation from
the intrinsic hydrogen column than does hard X-ray emission. Moreover,
as  the line  of sight  moves away  from the  projected  midplane, the
absorbing hydrogen column gets  smaller, resulting in smaller hardness
ratios.  It is therefore difficult to break the degeneracy between (a)
true spatial distribution of soft  vs.~hard X-ray emitting gas and (b)
preferred soft-end absorption by the BCG hydrogen column.
By iteratively increasing the magnitude of the photoelectric absorption 
component in a \texttt{WABS}$\times$\texttt{MEKAL} fit to the X-ray spectrum, we have estimated that a hydrogen column of $\gae10^{23}$ cm\mtwo\ would be 
required to attenuate enough of the soft X-ray flux to give rise to such confusion.
Such a high column density is not consistent with results from our 
X-ray spectral modeling, as discussed in the section above and shown in Table 1.  
We therefore consider it likely that the hard excess observed in the galaxy 
midplane is real.

\subsection{Radial profiles of gas properties}  

\label{section:deprojection}

We obtain radial profiles of temperature, density, pressure, and entropy by 
extracting spectra from both ObsIDs 7329 and 6934 using a
series of 20 concentric circular annuli with a  1\farcs5 inner radius,
3\farcs6 stacking increment, and 74\arcsec\ outer radius. 
Weighted  response matrices
were generated, and the background-subtracted spectra from each annulus were 
fit simultaneously in  \texttt{XSPEC} with absorbed \texttt{MEKAL} models (as before, $N_H$, $Z$,
and $z$ were  frozen to the appropriate values). 
The extracted spectra are an  
emission-weighted superposition of
properties along the line of sight extending through a 3D distribution of
thermal plasma. Projection effects are therefore important, however 
typical ``onion skin'' spectral deprojection techniques are not appropriate 
in this case as the inner 100 kpc of X-ray emission shows very large departures from 
spherical symmetry. We therefore choose to show only projected spectral profiles\footnote{As a test, we did attempt
spherical deprojection, and the results were very similar to our projected profiles.}.

In   Fig.~\ref{fig:deprojection}{\it   a} through \ref{fig:deprojection}{\it   d}    we  show   the   projected
temperature, electron density, pressure, and entropy profiles (respectively) out to 100 kpc. 
The latter three quantities are derived from the \texttt{MEKAL} normalisation (emission measure) using the equations 
listed in the Fig.~\ref{fig:deprojection} caption, as well as a simple volume assumption. See \citet{tremblay11} for more detail on how these profiles were obtained. We mark the location of the X-ray cavity network on each profile, 
and note a marginal steepening of the temperature gradient within this region.  
All of our profiles are consistent with past studies of both the early 18 ksec (good time) {\it Chandra} observation (ObsID 922, \citealt{mcnamara01,cavagnolo09}) as well as the {\it XMM-Newton} observation of A2597 \citep{morris05}. 
We plot independent consistency checks on our density and entropy profiles by using our Beta model fit to the 
surface brightness profile and the entropy parametrization from \citet{cavagnolo09}, respectively. 
The star formation threshold marked on the entropy profile is 
discussed in \citet{tremblay_cooling}.

\section{X-ray Spectral Maps}

In the left panel of Fig.~\ref{fig:full_tmap} we present the projected
X-ray temperature  map. 330 MHz, 1.3  GHz, and 8.4  GHz radio contours
are overlaid in black, green, and white, respectively. The right panel
shows the  same map,  smoothed with an  adaptive gaussian kernel  as a
viewing  aid.   We  overplot  the adaptively  smoothed  X-ray  surface
brightness  contours in  black.  The  color map  encodes  the best-fit
\texttt{MEKAL}  gas temperature ($kT$)  using the  scale shown  at the
right  of the  figure. Both  panels are  aligned, and  share  the same
$\sim60$ kpc  $\times 60$  kpc FOV centered  on the X-ray  centroid at
RA=23h 25m 19.75s, Dec =-12$^\circ$  07' 26.9'' (J2000). East is left,
north is up.

\begin{figure}
\plotone{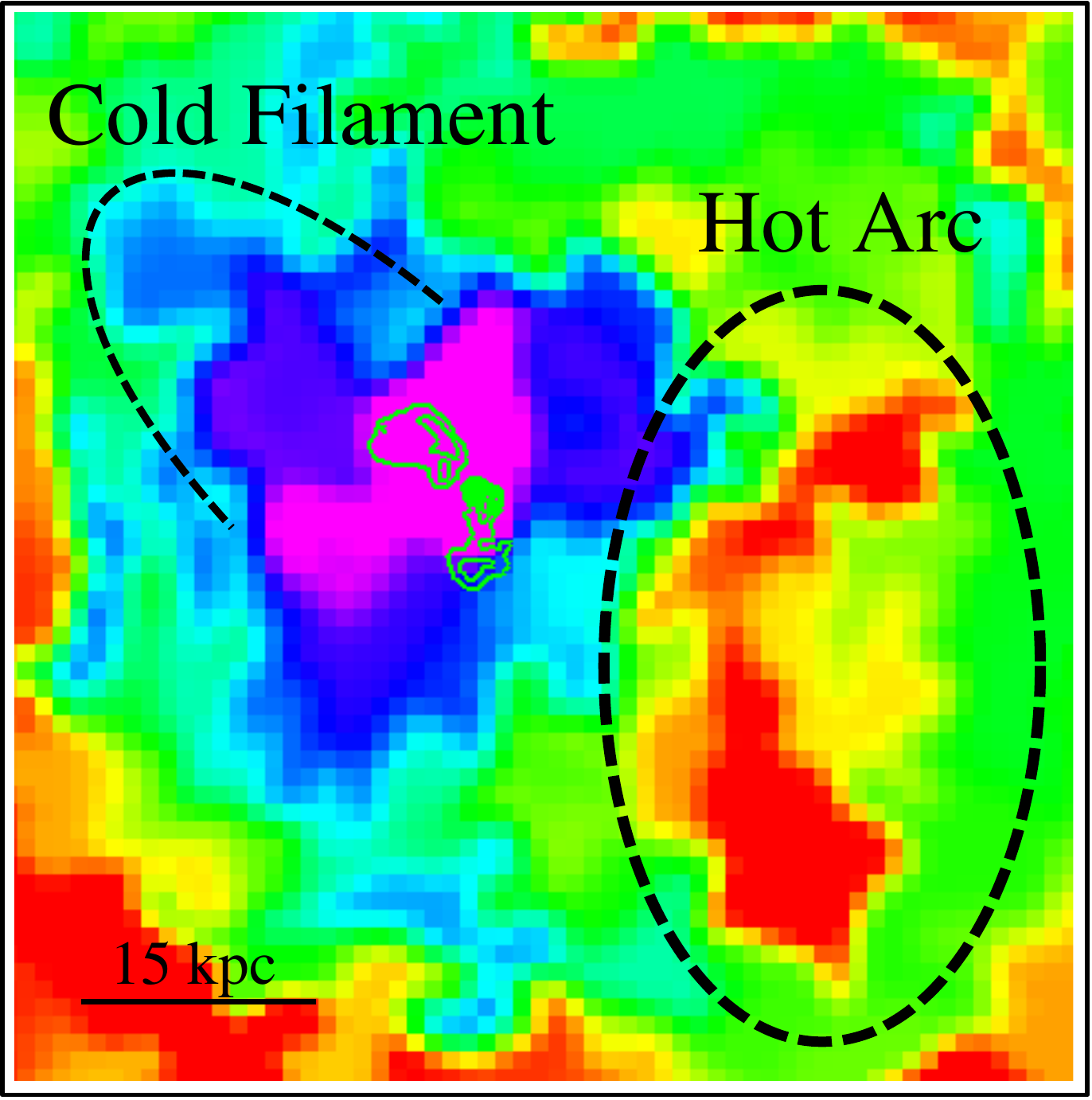}
\caption{Zoom-in on the X-ray temperature map shown in Fig.~\ref{fig:full_tmap}.  The central purple region of the map marks the location of the coolest ($\lae 2$ keV) X-ray gas. The cold filament and hot arc are labeled and outlined in black. Physical interpretations for these two features will be discussed in Sections 5 and 6.}
\label{fig:tmapfeatures}
\end{figure}

We  first note  that the  overall temperature  gradient  is reasonably
consistent with the projected  radial temperature profile (Fig.~\ref{fig:deprojection}{\it   a})  
over  the  same
radius.   We observe  two  significant spatial  fluctuations from  the
azimuthally   averaged  temperature   gradient,   namely  the   ``cold
filament'' and  ``hot arc'' features,  which are outlined  and labeled
accordingly in Fig.~\ref{fig:tmapfeatures}.

In comparing the top left panel with Fig.~\ref{fig:unsharp}, it can be
seen  that the  opening of  the hot  arc borders,  in  projection, the
eastern  edge of the  western large  cavity (features  1 and  2).  The
central axis of the extended western arm of the 330 MHz radio emission
spatially  coincides  with  the   projected  center  of  the  hot  arc
structure, just as it does the western cavity.

As  can be seen  by comparing  the temperature  map with  the overlaid
X-ray    surface    brightness    contours    (right    panel),    the
10\arcsec\ ($\sim15$  kpc) cold filament is nearly  cospatial with the
bright X-ray filament seen  in Fig.~\ref{fig:chandra}. In the hardness
analysis  presented in  Section~\ref{section:hardness},  we show  that
this  bright filament  is associated  with an  X-ray soft  excess (see
Fig.~\ref{fig:hardness}).   There is  slight spatial  mismatch between
the features, but this is not surprising given the spatial binning and
smoothing   processes  undertaken   during  creation   of   the  maps.
Inevitably, this  results in  some loss of  spatial information  on at
least  the  scale  of  the  contour bin  sizes.  But  when  considered
together,  Figs.~\ref{fig:hardness}  and  \ref{fig:full_tmap}  clearly
show  that   feature  (5)   in  Fig.~\ref{fig:unsharp}  is   a  ``cold''
filament.

\begin{figure*}
\begin{center}
\includegraphics[scale=0.51]{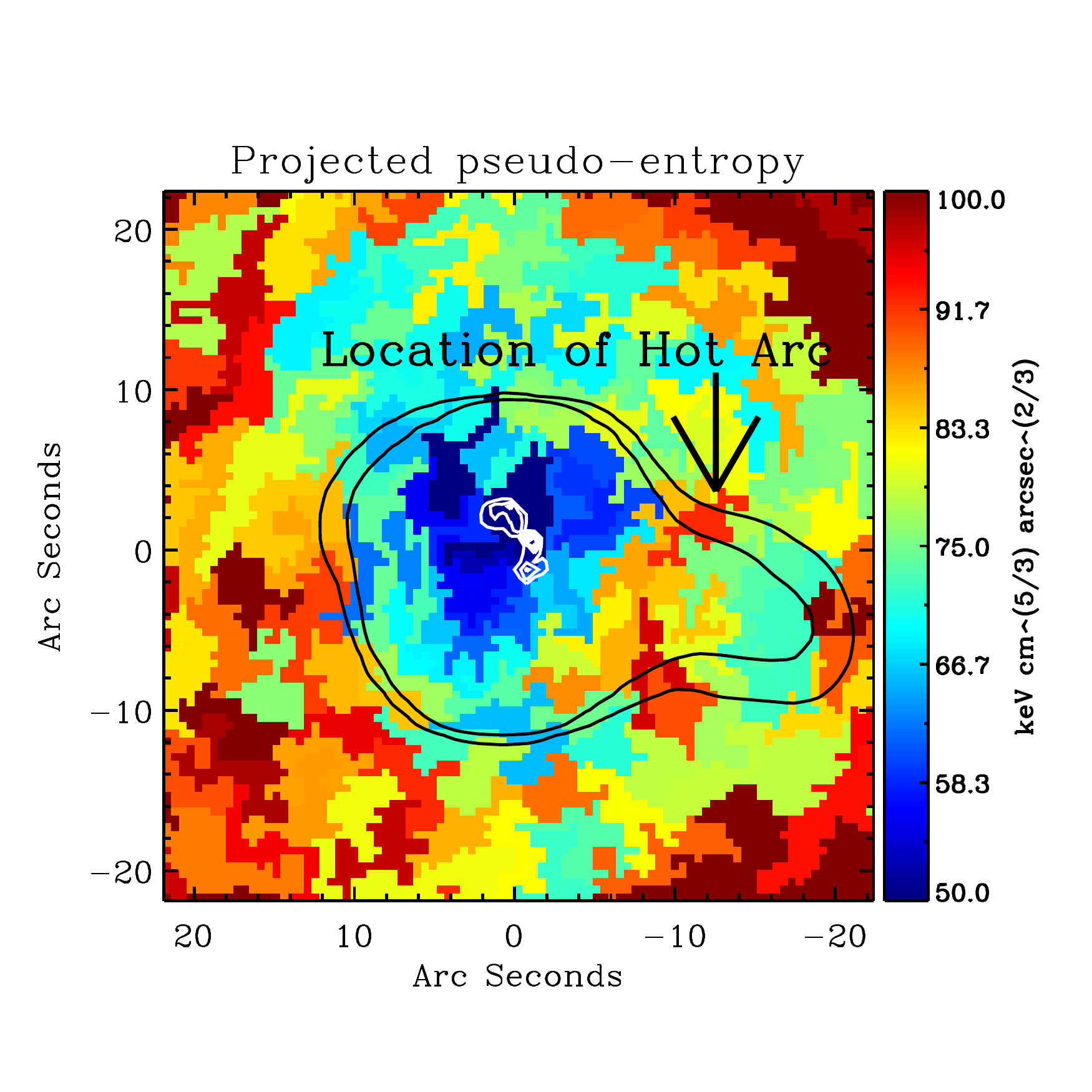} \hspace*{-6mm}  \includegraphics[scale=0.51]{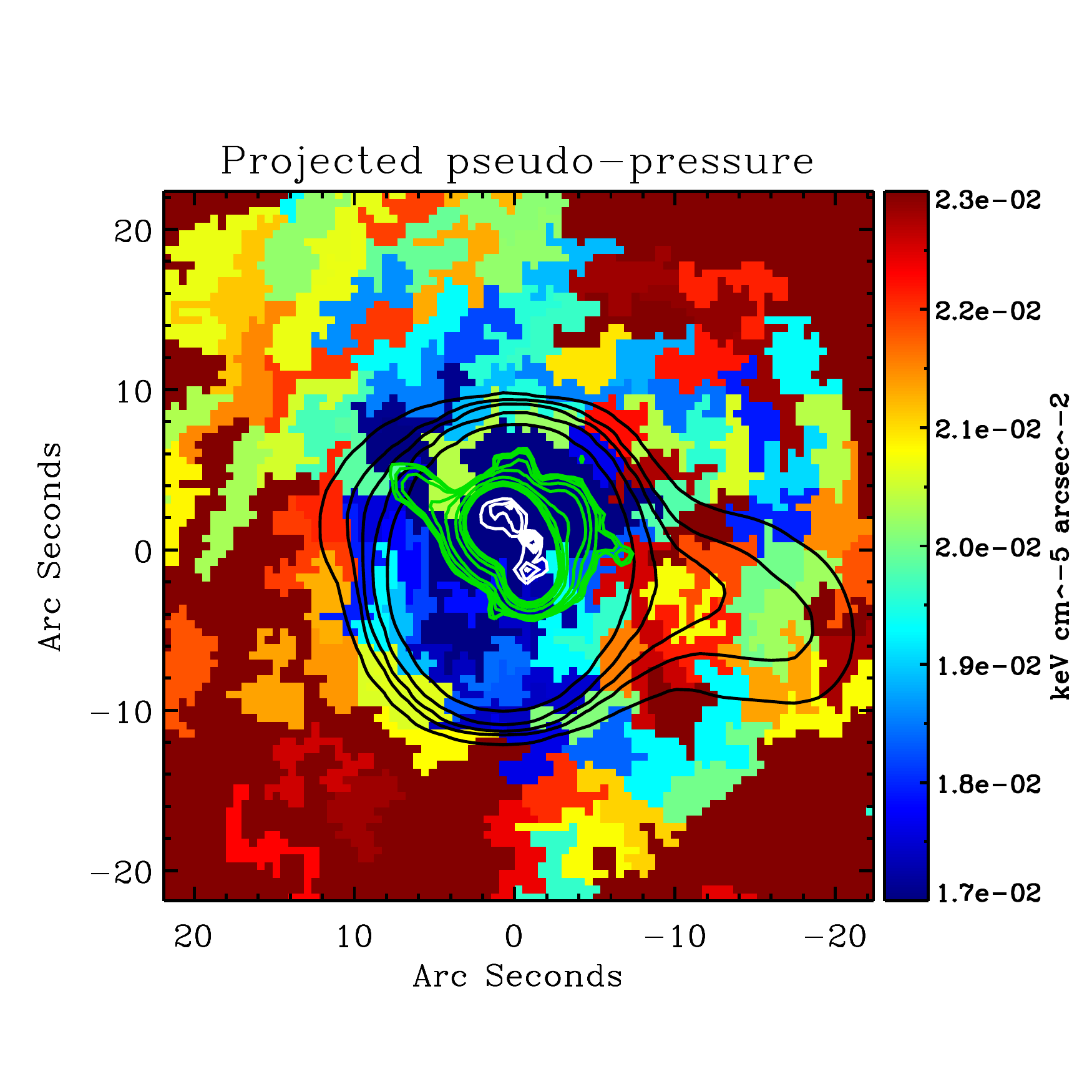}    \\
\vspace*{-12mm}
\includegraphics[scale=0.50]{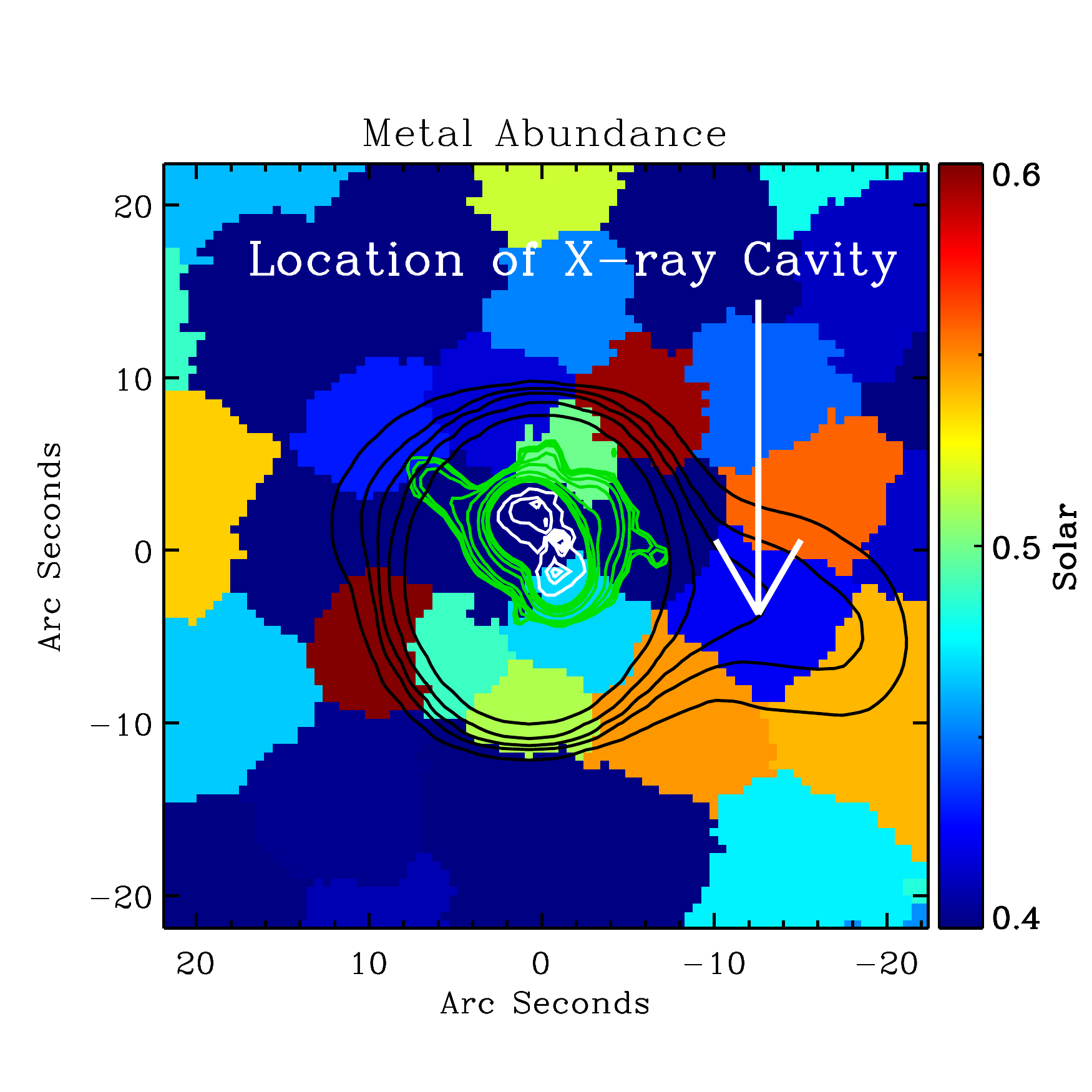}\includegraphics[scale=0.5]{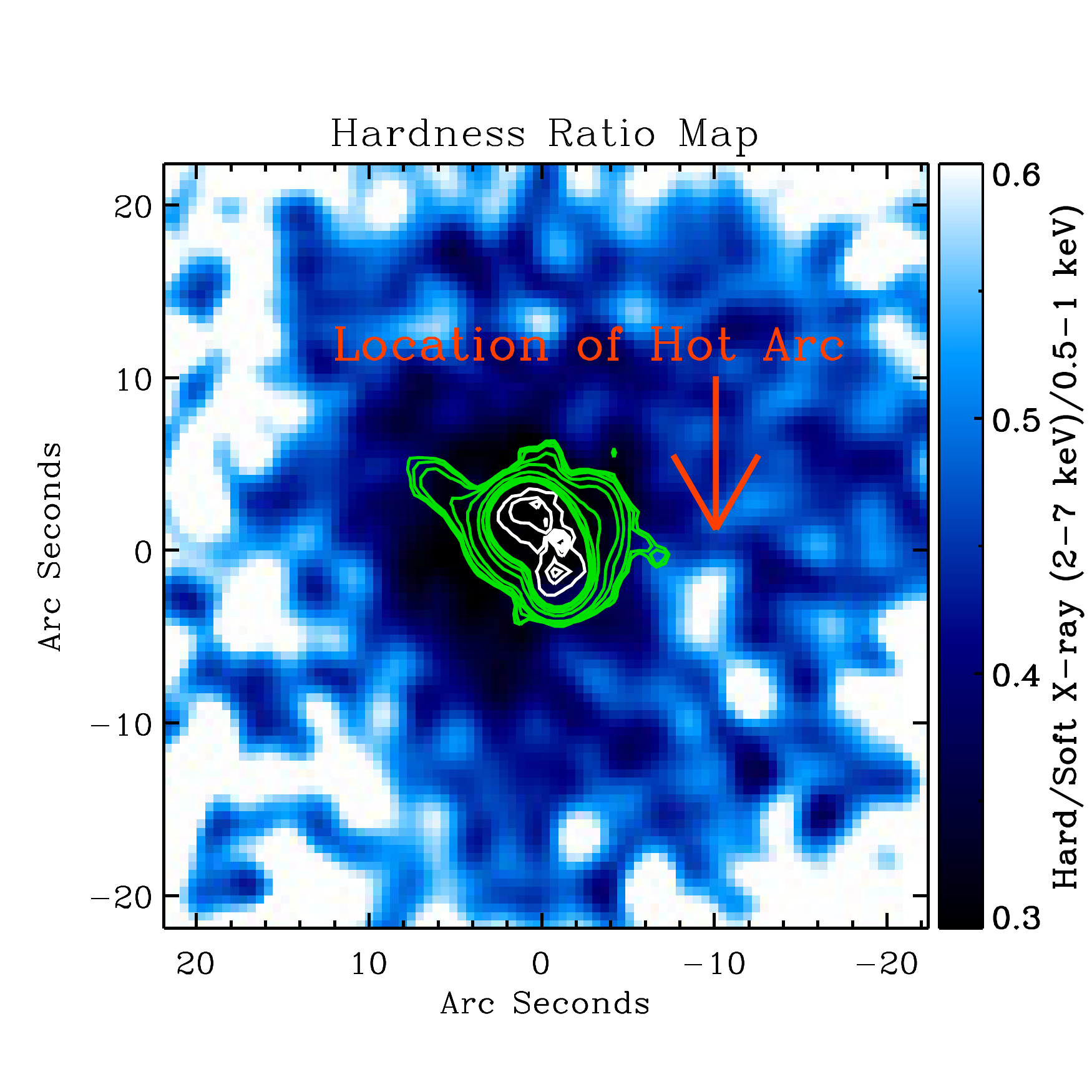}
\end{center}
\vspace*{-8mm}
\caption{({\it  top left})  The same  X-ray temperature  map  shown in
  Fig.~\ref{fig:full_tmap}, with  330 MHz, 1.3 GHz, and  8.4 GHz radio
  emission overlaid in black, green, and white contours, respectively;
  ({\it  top right})  Pseudo-pressure  map made  by  multiplying the
  temperature ($kT$)  map at  {\it left} with  the square root  of the
  emission measure map; ({\it  bottom left}) metal abundance map, made
  with larger  contour bins  set by a  S/N=70 threshold;  ({\it bottom
    right}) X-ray  hardness ratio map,  made by dividing  the exposure
  corrected 2-7 keV and 0.5-1 keV  data. A gaussian kernel was used to
  smooth the hardness ratio map to roughly the same spatial resolution
  of the  X-ray temperature and  pressure maps.  All panels  share the
  same FOV, and are roughly 60 kpc  on a side. The location of the hot
  arc (or  large western cavity in  the case of the  abundance map) is
  marked with an arrow. }
\label{fig:temppressure}
\end{figure*}

In Fig.~\ref{fig:temppressure}  we show the pseudo-entropy\footnote{We
  use  the term ``pseudo''  because  the square  root  of the  emission
  measure (i.e., the \texttt{MEKAL}  normalization) is used  as a proxy  for the
  gas density.  No inherently uncertain volume assumption  is made, so
  this   cannot   be  considered   the   actual   gas   density.},
pseudo-pressure, metal abundance, and hardness ratio maps.  We overlay
330  MHz and  1.3 GHz  contours on  the pseudo-entropy  map  (top left
panel) in black and white, respectively. 330 MHz, 1.3 GHz, and 8.4 GHz
contours are  shown on the  pseudo-pressure map (top right)  in black,
green, and  white, respectively.  We intentionally show  fewer 330 MHz
contours on the pseudo-entropy map to  allow the hot arc feature to be
clearly visible. On each map, the location of the hot arc feature seen
in  the  X-ray temperature  map  is  marked.  Both pseudo-entropy  and
pseudo-pressure are elevated in the  hot arc region.  This result will
be discussed in more detail in Section 6.

In the  bottom left panel  of Fig.~\ref{fig:temppressure} we  show the
metal abundance map.  \citet{kirkpatrick11} studied metal abundance in
A2597  in  more detail  than  we will  here.  That  work found  higher
abundance gas extended  along the cavity and radio  axis in A2597, and
suggested  that this could  be evidence  for a  metal-enriched outflow
driven  by  the  radio  source.   Such a  result  is  consistent  with
predictions      from     theoretical     AGN      feedback     models
\citep[e.g.,][]{pope10,gaspari11},  as well as  results from  other CC
clusters   (e.g.~Hydra    A,   \citealt{simionescu09,gitti11}).    Our
abundance map  has been  made with large  spatial bins using  a S/N=70
threshold,  as noted  previously.   This results  in  loss of  spatial
information on the  scale of the cold filament and  hot arc, though we
do  note that  higher abundance  gas is  generally extended  along the
cavity   and   radio   axis,   consistent  with   the   results   from
\citet{kirkpatrick11}.  Higher-than-average  abundance is not observed
at the  location of  the western large  cavity (marked with  the white
arrow), but it is seen along  its edges. This might be expected if the
metal-enriched gas was pushed outward along the bubble boundary during
the cavity excavation process.

In Fig.~\ref{fig:temppressure}{\it d} we  show an X-ray hardness ratio
map,  made  by  dividing the  hard  energy  slice  by the  soft,  then
smoothing with  a 1\farcs5 Gaussian kernel  (non-adaptive). The kernel
size was chosen to roughly  correspond with the spatial bin sizes used
in the creation  of the 2D spectral maps. The location  of the hot arc
corresponds to  a region of relative  hard excess that  is roughly the
same shape.  Some mis-match  in shape is  expected, because  the X-ray
temperature map was created  with the spatial contour binning process,
and  the  hardness  ratio  map   was  not.  As  discussed  in  Section
\ref{section:hardness}, there  is always some degeneracy  in this type
of hardness analysis, as a region of hard excess can't necessarily
be  distinguished  from  a  region  of  soft  deficit  or  locally
higher-than-average  soft-end absorption  by  a more  dense column  of
intervening hydrogen  along the line  of sight. In principle,  the hot
arc could merely be an artifact of this degeneracy, or a superposition
of otherwise unrelated foreground features.

We find  these latter possibilities unlikely, simply  because it's not
clear  why the  soft X-ray  extinction would  be that  patchy  on such
spatially small scales. It would also be a remarkable coincidence that
a  superposition  of foreground  features  just  happens  to align  in
projection  with the  inner  edge  of the  western  X-ray cavity.  In
Section 6,  we discuss how  the hot arc  may, given a large  number of
strong caveats, be a signature of ICM/ISM heating by the western large
cavity  as it  buoyantly rises,  dissipating its  enthalpy as  heat as
ambient gas moves to refill its wake.

\section{Physical Interpretation of the Cold Filament}

\label{section:dredge}

As can be seen in Fig.~\ref{fig:full_tmap}, the 1.3 GHz radio emission
(green contours)  features a northeastern hook aligned  along the same
position  angle as  the  cold  filament. While  the  extended 1.3  GHz
emission in this area covers only a small spatial fraction of the cold
filament,  it could  be interpreted  as evidence  of dredge-up  of low
temperature, high  density X-ray gas by the  propagating radio source.
Both the cold filament and the 1.3 GHz hook are aligned along the same
position  angle as  the western  large  cavity, the  extended 330  MHz
emission, the projected VLBA jet axis, and the high velocity stream of
warm optical  line emitting gas  observed in the  \citet{oonk10} data.
Although the  1.3 GHz hook is  only cospatial with the  bottom half of
the cold filament,  this should not necessarily be  interpreted as the
terminus of  the jet.  The  1.3 GHz data  may simply not  be sensitive
enough  to sample  the regions  of the  jet beyond  this  location, or
synchrotron  losses   could  have  rendered  this  part   of  the  jet
unobservable at  these frequencies.  The eastern extended  ``bump'' of
330 MHz  emission (black contours) suggests that  radio structures lie
beyond the maximal cluster-centric radius of the 1.3 GHz emission.

\subsection{Possible theoretical models}

\begin{figure*}
\includegraphics[scale=0.6]{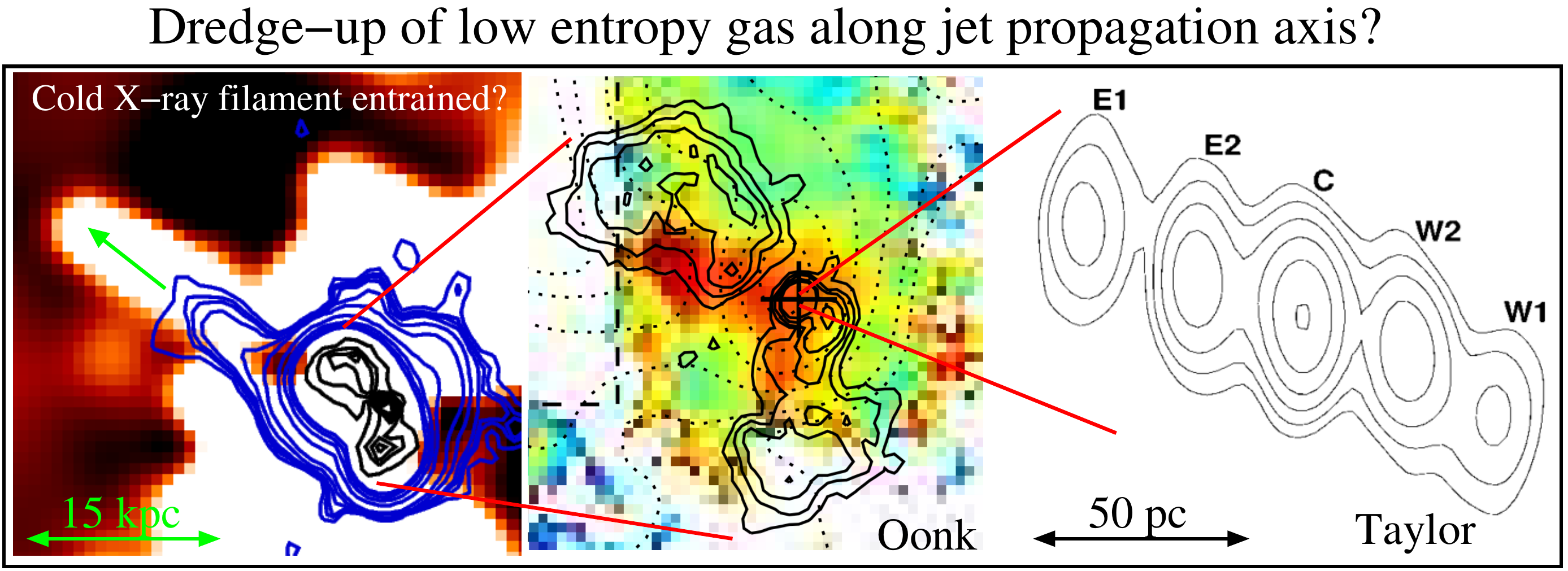}
\caption{In the left panel, we show 8.4 GHz radio emission, in black, and 1.3 GHz radio emission,
  in   blue,    overlaid   on   the   residual    X-ray   image   from
  Fig.~\ref{fig:unsharp}. The eastern edge  of the 1.3 GHz emission is
  extended along the  bottom half of the cold  X-ray filament. The
  western edge  of the 1.3 GHz  emission is extended  in the direction
  along  the western  large cavity,  along with  the 330  MHz extended
  emission, suggestive  of a common  axis offset from the  central 8.4
  GHz source. That the 1.3 GHz  hook is cospatial with the soft excess
  X-ray filament may be consistent with a scenario wherein low entropy
  gas is  dredged up  by the propagating  radio source  (either during
  this  epoch of activity  or a  previous one),  as has  recently been
  observed  in Hydra  A \citep{gitti11}.  The  center panel  is a  VLT
  SINFONI Paschen-$\alpha$ velocity dispersion map on the scale of the
  8.4 GHz source, from  \citet{oonk10}. The higher velocity dispersion
  gas  (in red) is  aligned with  the projected  VLBA small  scale jet
  axis,    which   we    show   in    the   right-most    panel   from
  \citet{taylor99}. This  could be  evidence for entrainment  of dense
  material  by  the  propagating  radio  source. The  FOV  is  of  the
  left-most panel  is approximately $\sim 30$ kpc  $\times30$ kpc. The
  panels subsequently zoom  in from there, to $10$  kpc $\times10$ kpc
  and $100$ pc $\times100$ pc, respectively. }
\label{fig:filament}
\end{figure*}

We consider four possibilities for the cold filament.

\begin{enumerate}

\item The cold filament could simply be the border between the neighboring cavities (features 3 and 4 Fig.~\ref{fig:unsharp}).

\item The cold filament could be due to a morphologically complex multiphase cooling region around 
the bubbles.

\item Similar to scenario (ii), the filament could arise from displaced keV gas
funneling (but not necessarily cooling) down the region separating the two neighboring X-ray cavities (features 3 and 4), or could arise from 
one larger cavity disrupting into two, allowing for gas flow down what 
was formerly the center of the larger cavity.

\item The cold filament may be associated with dredge-up of multiphase 
gas by the propagating radio source and dragged outwards. The main 
evidence for this is that the filament is aligned along the same position angle
as the projected jet axis, and partially cospatial with with the extended 1.3 GHz hook.  
As we will discuss, there is additional multiwavelength evidence in support of this scenario. 

\end{enumerate}

We first note that our ability to discriminate between these scenarios is strongly limited by unknown 
projection effects. With this caveat in mind, we note that   
possibility (i), wherein the filament may simply be the border between two cavities, is the simplest and perhaps the most likely explanation as it 
relies on the least number of assumptions. This would not necessarily explain the observed 
soft excess associated with the filament.

Possibility (ii), wherein gas cools in a spatially complex manner around the cavities, 
is consistent with inhomogeneous cooling flow models generating thermally unstable filaments (e.g., \citealt{sharma11,fabian11}), and could explain the multiwavelength results discussed in \citet{tremblay_cooling}.
As discussed below, the cold filament is adjacent to (and not cospatial with) the optical emission line filaments 
shown in Fig.~6 of \citet{tremblay_cooling}. This is contrary to simple expectations if the filament represents the hot phase of a multiphase cooling region that is forming the optical emission line filaments. Instead, the spatial relationship between the X-ray and optical filaments seems more consistent with scenarios in which the 
optical filaments have been swept aside by the 15 kpc X-ray filament, rather than formed from it.

Possibility (iii) is similarly difficult to interpret quantitatively, as
projection effects make it impossible  to know the real spatial relationship
between the filament and its two apparently neighboring cavities.  
If   two X-ray bubbles were buoyantly
rising side-by-side, a large volume of the displaced ambient keV gas could be
preferentially funneled into a column separating the two cavities. A larger 
bubble which has disrupted may result in a similar effect.
Again, it is difficult to be more quantitative in considering these scenarios, 
for the reasons discussed above.

While much  of this discussion  is speculative, we note  that possibility
(iv)  is supported  by  additional observational  evidence.  In  this
scenario,  the  cold  filament  may  arise  from  dredge-up  of  lower
temperature, higher density X-ray gas by the propagating radio source,
as   has  recently   been  observed   in  the   CC  cluster   Hydra  A
\citep{gitti11}. We provide a schematic  for this scenario and some of
this  supporting evidence  in Fig.~\ref{fig:filament}.   Comparing the
leftmost and rightmost panels of this figure, note the close projected
position  angle  alignment  between  the  cold filament  and  (1)  the
extended ``hook''  of the 1.3  GHz radio emission (blue  contours) and
(2) the  projected current VLBA  jet axis \citep{taylor99}.  Note also
that the filament is extended along the 330 MHz radio and X-ray cavity
axis  (see  Fig.~\ref{fig:unsharp}).   Furthermore,  the  X-ray  metal
abundance  map  shown   in  Fig.~\ref{fig:temppressure}{\it  c}  shows
evidence for higher metallicity keV gas extended along this same axis,
consistent with a  picture in which enriched gas  from the BCG nucleus
has been dragged outwards.

Meanwhile, there have been  prior suggestions that the ``current'' jet
associated  with the  AGN  may  be entraining  ambient  warm and  cold
molecular gas.  \citet{oonk10} presented $K$-band integral field (IFU)
spectroscopy  enabling gas  kinematics analysis  of the  molecular and
ionized gas distribution on the scale  of the A2597 nebula and the 8.4
GHz radio source. They  reported high velocity and velocity dispersion
($\sim200-300$   km  s$^{-1}$)  streams   of  H$_2$   and  \ion{H}{ii}
coincident with the  southern edge of the northern  8.4 GHz radio lobe
and  approximately  aligned  with  the  VLBA jet  axis.  Another  high
velocity  dispersion filament  is  coincident in  projection with  the
eastern edge of  the southern lobe.  Their velocity  dispersion map is
shown in the  center panel of Fig.~\ref{fig:filament}.  \citet{oonk10}
considered   two  possible  explanations   for  these   high  velocity
dispersion features. If  the radio source is a  wide angle tail (WAT),
the high velocity  filaments may arise from the  turbulent wake caused
by  relative motion of  the AGN  amid the  ambient dense  medium. This
seems unlikely as the  source is uncharacteristically compact relative
to  known WAT  sources, and  probably  too small  to be  significantly
affected by the host galaxy motion through the ICM.

Alternatively, the  close projected alignment of the  current VLBA jet
axis  for  the  northern  lobe  is suggestive  of  direct  kinematical
interaction  (e.g., mass  entrainment)  between the  jet  and the  gas
through which it is propagating.  In the latter case, the symmetry and
asymmetry  of the VLBA  and 8.4  GHz counterjets,  respectively, would
require significant deflection  of the current jet to  account for the
position angle mismatch  between the 8.4 GHz lobes.  A gradual bend in
the jet owing to relative motion (consistent with the former scenario)
may account  for this.  Interaction with the  ambient medium  may also
produce  a  sharp  trajectory  deflection of  a  rapidly  decelerating
counterjet. The site of this possible deflection is not visible in the
\citet{oonk10} velocity maps, though the  bright knot of 8.4 GHz radio
emission immediately  SW of the  core may be related.   Both scenarios
are  consistent  with the  steep  lobe  spectral  index suggestive  of
dynamical frustration and confinement.

The bottom  half of the cold  filament is cospatial  with extended 1.3
GHz radio emission (Fig.~\ref{fig:filament}, blue contours). The width
of  this extended  hook is  approximately  the width  of the  filament
($\sim 2$  kpc), though it  is slightly offset  to the south  from the
filament's major axis. Of course,  the 1.3 GHz emission may not sample
the entire jet, regardless of  whether or not the filament was dragged
upwards  during this or  a previous  epoch of  activity.  The  1.3 GHz
emission may  sample only a small  portion of the real  jet, while the
remainder of  the emission  may have aged  to lower  frequencies.  The
lower frequency 330  MHz emission is not resolved  on these scales, so
it is difficult  to determine whether this may be  the case, though it
is worth noting  that the 330 MHz emission does  show a slight eastern
extension at the same radius as  the terminus of the cold filament, so
some part  of the jet  (relic or otherwise)  has made it out  to these
radii.  Deeper  radio data  are needed to  further study  the apparent
connection to the filament.

\begin{figure*}
\begin{center}
\includegraphics[scale=0.5]{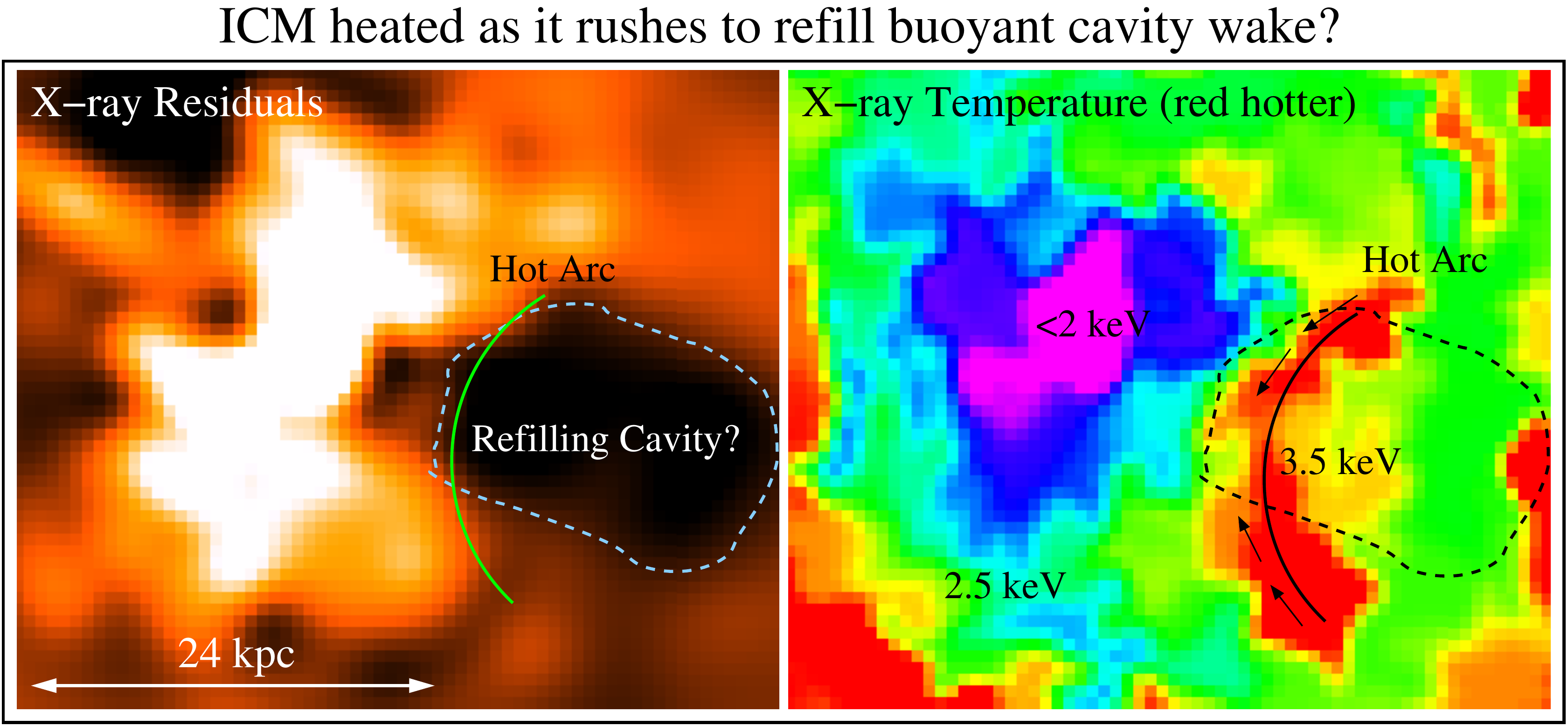}
\end{center}
\caption{Zoom-in on ({\it left}) the residual X-ray image presented in
  Fig.~\ref{fig:unsharp}  and  ({\it right})  the  temperature map  to
  highlight the temperature features associated with the western large
  cavity, which we  highlight with a dashed line  in both figures. The
  FOV of both panels is identical. The green and black arcs traces the
  ``hot arc'' feature previously described in the temperature map.  In
  the  temperature map  we  draw arrows  that  illustrate our  working
  interpretation for the ``hot  arc'' feature. In this interpretation,
  the keV  gas displaced by the  buoyant rise of the  M01 ghost cavity
  rushes  inward to  refill  the resulting  wake, thermalizing  cavity
  enthalpy  by heating  the ambient  gas  $>1$ keV  above its  ambient
  surroundings.  }
\label{fig:hotarc}
\end{figure*}

\subsection{Energy arguments for the dredge-up model}

As recently done in \citet{gitti11} for Hydra A, we can roughly estimate the
energy that would be required to lift cold gas from the core to the projected
height of the filament. To first order, this will be the difference in
gravitational potential energy between the two locations.  Assuming the local
hot ISM is isothermal, hydrostatic, and with a sound speed
$c_s \approx 750$ km s$^{-1}$ \citep{tremblay_cooling},
this will be 
\begin{equation} 
\Delta E = \frac{M_\mathrm{gas} ~c_s^2}{\gamma}
\ln \left(\frac{n_{e,i}}{n_{e,f}}\right).  
\label{eqn:gitti}
\end{equation}
 where $\gamma\simeq5/3$  is the ratio of specific  heats.  We roughly
 estimate the mass  of the displaced X-ray gas (e.g.,  the mass of the
 cold  filament),  $M_\mathrm{gas}$, by  scaling  inferred values  for
 X-ray gas mass within  $\sim 100$ kpc to a radius of  15 kpc, 
then scaling further to  account for the filling factor of
 the  cold filament.  Allowing  for a  conservative range  of inferred
 X-ray masses  and filling factors,  we estimate a cold  filament mass
 range of $10^7-10^9$ \Msol.  The gas density at the bottom and top of
 the filament  ($n_{e,i}$ and $n_{e,f}$,  respectively) were estimated
 from the radial density profile (Fig.~\ref{fig:deprojection}).  We find the
 energy required to lift a $10^7$  \Msol\ filament out to 15 kpc to be
 $5.4\times10^{56}$  ergs,  while the  energy  required  for a  10$^9$
 \Msol\ filament is  on the order of $10^{58}$  ergs. These values are
 comparable to the $pV$ energy of the X-ray cavities listed in Table 2
 of \citet{tremblay_cooling}, which can be considered  rough lower limits on the AGN power
 output. If  the AGN has  been powerful enough  in the past  to expend
 $>10^{56}$  ergs   to  excavate  kpc-scale  X-ray   cavities,  it  is
 energetically feasible that it could  also expend a similar amount of
 energy to entrain and lift cool X-ray gas out to 15 kpc.

\section{Physical Interpretations of the Hot Arc}

We consider the following possibilities for the hot arc feature.

\begin{enumerate}

\item It could be an artifact of the process employed to make the spectral maps and not associated with any 
real feature; or the spatial binning process could have exaggerated the 
significance and strength of a generally warmer but otherwise unremarkable region;

\item The hot arc could be a superposition of real foreground 
features, or a localized region of soft X-ray {\it deficit} rather than a 
true hard X-ray {\it excess} that would be associated with truly hotter gas;

\item The feature could be an artifact caused by a cool outer arc
pushed out by the 330 MHz bubble; 

\item  The hot arc could be a signature of the cavity heating process.

\end{enumerate}

Concerning  possibility (i):  We have  made numerous  versions  of the
temperature map (for example,  with various S/N binning thresholds and
bin size ratio  constraints) in an effort to  explore the significance
of  the hot  arc result.   In the  temperature map  presented  in this
paper, there are $\sim14$ (S/N=30, $\gae 900$ count) bins that make up
the hot arc feature, depending on how one defines its edges.  Adjacent
bins are fit with consistent temperatures that follow a steep gradient
to the ambient colder temperature.  The errors on the temperature fits
to the spatial bins in the hot arc region are $\pm 0.25$ keV.  In test
maps made with many small (S/N=10) bins, the feature persists. In maps
made  with   very  large  round  bins  (e.g.,   S/N=70),  the  feature
disappears,  though   this  is  not  surprising.    How  the  spectral
extraction bins are tiled over the  hot arc region has a strong effect
on the  apparent significance  and shape of  any such feature.  A real
sharp-edged localized  spike in temperature  could be smoothed  out or
``smudged'' by bins which span both sides of the feature, resulting in
temperature fits  that reflect  an average of  the hotter gas  and the
ambient colder gas.

Concerning caveat  (ii), we  note that the  feature does appear  to be
associated with an arc in the 1-2 keV panel of Fig.~\ref{fig:hardness} (which we point out with an arrow), though its shape is somewhat dissimilar (this could simply be a consequence of smoothing). The hot arc is also associated with a hard excess  arc in the hardness ratio map (Fig.~\ref{fig:temppressure}{\it d}).  Our
X-ray observations are  not deep enough, lacking the  counts needed to
determine whether the  arc may be a site  of higher intrinsic hydrogen
column  density  that  would  lead  to  more  soft  X-ray  absorption,
especially  considering  the   degenerate  nature  of  X-ray  spectral
fitting.  Given the strong spatial  correspondence of the arc with the
location and linear  extent of the western cavity,  the orientation of
the 330 MHz radio source, VLBA  jet, and 1.3 GHz emission, we find the
latter  possibility   unlikely  (because  it  would  be   a  very  big
coincidence). For this  same reason, we find it  unlikely that the arc
is a superposition of foreground features.

Assuming the hot arc is indeed a real feature, we speculate below 
on two possible physical interpretations (i.e., scenarios iii and iv).

\subsection{Cool gas pushed outwards by the radio bubble?}

The  hot arc  may be  part of  the general  temperature  gradient, but
appear isolated because  the 330 MHz radio bubble  associated with the
western large cavity  has compressed and uplifted a  rim of cooler gas
from the center.  Entrainment and displacement of colder ISM phases by
ascending X-ray cavities  has been observed before in  other cool core
clusters (e.g., recently  in Perseus/NGC 1275, \citealt{fabian11}; and
M87,  \citealt{werner10}).   Heightened   metal  abundance  along  the
cavity/radio  axis (see  Fig.~\ref{fig:temppressure}{\it  c}) supports
the  possibility  that  the  propagating radio  source  has  displaced
enriched colder gas  from the core. The energy  arguments for the cold
X-ray filament discussed in Section 5  still apply here --- the AGN is
almost certainly powerful enough to  uplift a significant gas mass out
to  a radius  of $\sim  20$  kpc. The  radio source at various  frequencies exhibits evidence
for complex  dynamical interaction with  optical and FUV  filaments on
$\sim 10$  kpc scales (e.g., \citealt{tremblay_cooling}). It  seems reasonable that  similar interactions
could be occurring at $\sim 20$ kpc scales.  The cooler region seen to
the West of the hot arc  ends at the projected terminus of the western
large cavity. Beyond it, the  spectral map shows hotter $\sim 3.5$ keV
temperatures consistent with the  local radial average.  Uplifted cool
gas is expected to have important consequences as it mixes with hotter
gas at  larger radii, such  as lowering the  net mass inflow  rate and
giving rise to efficient  thermal conduction and ionizing interactions
at the mixing interface (e.g., \citealt{werner10}).

\subsection{A signature of cavity heating?}

In  Fig.~\ref{fig:hotarc} we  provide a  schematic for  an alternative
interpretation  of  the hot  arc,  assuming  it  is real.   In  simple
AGN-driven  cavity heating  models, the  enthalpy (free  energy)  of a
buoyantly rising  cavity can be  entirely dissipated as  the displaced
thermal  gas  rushes to  refill  its wake  (see  e.g.,  the review  by
\citealt{mcnamara07}).  As the gas moves  inward to fill the void left
by  the bubble,  the corresponding  change of  gravitational potential
energy  is  converted to  kinetic  energy,  which  then dissipates  as
heat. In principle,  the kinetic energy created in  the cavity wake is
exactly  equal  to  the  enthalpy   lost  by  the  cavity  during  its
corresponding  ascent.   Regardless  of  the Reynolds  number  of  the
plasma, viscous dissipation and turbulence ensures that kinetic energy
of  the infalling  gas  is damped  rapidly,  so thermalization  occurs
directly behind and on the same general scales as the bubble.

As seen in  Fig.~\ref{fig:hotarc}, the hot arc borders  the inner edge
of the western  cavity, so its shape and  location are consistent with
the above scenario in which the  gas in the arc has been heated during
the buoyant passage of the bubble. We can make two simple estimates to
determine whether this makes sense  in the context of available energy
budgets. We  first roughly estimate the total  enthalpy $H$ associated
with the  western cavity, which is the  sum of the $p  \times dV$ work
associated with inflation  and the thermal energy $E$  of the cavity's
contents:
\begin{equation} H = E + pV =
\frac{\gamma}{\gamma-1} pV, 
\end{equation} where $V$ is the cavity volume and $p$
is  the  pressure  of  the  radio lobe  which  displaced  the  thermal
gas. Depending  on its  contents, cavity enthalpy  is in the  range of
$2pV - 4pV$. Using this,  we estimate the enthalpy associated with the
western large cavity to be  $\sim7\times10^{58}$ erg, which is a rough
limit  on the  energy reservoir  that is  (in principle)  available to
eventually heat the ICM over  some unknown timescale (that is probably
limited by the cavity lifetime).

As we did  for the cold filament, we  can use Equation~\ref{eqn:gitti}
to estimate  the change  in gravitational potential  energy associated
with displacement  of the  hot arc gas.   Using the same  (very rough)
estimation strategy as  before, we assume that the  mass of gas making
up the  hot arc  is in the  range of  $10^7-10^9$ \Msol\ and  that the
density gradient  and sound speed is the  same as it was  for the cold
filament (the  two features are roughly at  similar radial distances).
In  this case, the  gravitational potential  energy change  is between
$10^{56} \lae  \Delta E \lae 10^{58}$  ergs, of order what  it was for
the  cold filament.   If  we assume  this  energy heats  the gas,  the
corresponding  temperature  change  can  be simply  approximated  with
$\Delta E= (3/2) N k \Delta  T$, where $N$ is the number of particles.
In estimating $N$, we use the electron density profile obtained in Section 3
and  assume  that  the   volume  of  the  hot  arc  can  be
approximated by a  spherical cap shape between 1\%-30\%  the volume of
the  western large  cavity.   We  find that,  in  principle, there  is
roughly enough energy to raise the local temperature of the hot arc by
$\sim 1$ keV, which is what is observed in the temperature map.

Obviously,  there  are  extremely  uncertain assumptions  behind  this
estimate,   so  this   is   only  useful   as  an   order-of-magnitude
consideration. In  principle the hot arc  would tap a  fraction of the
instantaneous  cavity  power  $P_\mathrm{cav}$  of the  western  large
cavity,  or about  $\sim10^{42}-10^{43}$ ergs  s$^{-1}$ \citep{tremblay_cooling}. 
The cooling luminosity associated with the inferred
classical  cooling  flow  is  on  the order  of  $3\times10^{43}  \lae
L_\mathrm{cool}  \lae  10^{44}$ ergs  s\mone,  so  the  hot arc  could
theoretically be associated with a significantly large fraction of ICM
radiative losses.   If real,  this feature could  be one of  the first
known observational signatures of the AGN cavity heating model invoked
at late epochs to quench cooling flows \citep{mcnamara07}.

If we assume  that the arc is real and  the interpretation is correct,
then why  is A2597 the only known  cluster with a hot  arc?  There are
numerous  deep {\it  Chandra}  observations of  X-ray  cavities in  CC
clusters with published temperature maps (including maps with many more counts per square pc than ours), 
and no such feature has ever
been  reported  before.  Shocks  and  cold  fronts  can  have  similar
morphologies as the hot arc,  but our results are inconsistent in both
orientation  and  spectral characteristics  with  these more  commonly
observed phenomenon. If  the hot arc were a  shock associated with the
radio source,  for example, we might  expect the feature to  be at the
outer edge  of the  radio bubble with  a concave-inward  shape.  There
could be  serendipity associated  with detection of  this feature.
If we assume that all buoyantly rising X-ray cavities produce hot arcs
in  their   wakes,  their  detectability  would   likely  require  (1)
sufficiently  deep X-ray  observations; (2)  X-ray spectral  maps made
with spatial bins  that are individually smaller than  the cavity; and
(3) a large, $\gae 10$ kpc  cavity that is buoyantly rising in roughly
the  plane of  the  sky.  The latter  requirement  seems necessary  to
recover  the arc  shape and  concave-outward orientation.   All of
these  detectability  requirements  are  further  dependent  upon  the
unknown lifetime  of such a  temperature feature, especially if  it is
short lived.  One possible explanation for  why we do not see hot arcs
associated with  other cavities is  that the locally heated  gas mixes
with the ambient colder gas or  cools very quickly. The lifetime of an
arc would  strongly depend on the  unknown role played  by gas motions
and/or thermal conduction.

\section{Summary}

The primary results of this paper can be summarized as follows. 

\begin{itemize}

\item New {\it Chandra} X-ray observations of Abell 2597 reveal a highly 
anisotropic surface brightness distribution, permeated by a network 
of X-ray cavities that is more extensive than previously known. The largest cavity 
is cospatial with extended 330 MHz radio emission. 

\item A $\sim 15$ kpc soft excess X-ray filament is found along the 
same position angle as the cavity/radio axis, and is partly cospatial 
with a hook of extended 1.3 GHz radio emission. Among several possible scenarios, 
we discuss multiwavelength evidence suggesting that the filament 
could be associated with the dredge-up of multiphase ($10^3-10^7$ K) gas 
by the propagating radio source. Although other interpretations are possible, 
we note that this model would be one of the most 
dramatic known examples of such an interaction. 

\item X-ray spectral maps reveal an arc of hot, high entropy gas
bordering the inner edge of the largest X-ray cavity. We suggest that, 
if the feature is not an artifact, it could be due to (1) an uplifted rim of 
cold gas from the core, pushed outwards by the radio bubble; or (2) associated
with cavity heating models invoked to quench cooling flows within the 
radio-mode AGN feedback interaction region. 

\end{itemize}

Deeper radio data will be needed to study 
multiphase gas entrainment along the filament, and deeper 
X-ray data are needed to further understand the significance and implications of the hot arc.

\section*{Acknowledgments} The authors thank Drs. Elaine Sadler, Robert Laing, 
Andy Robinson, Joel Kastner, and Bill Sparks for thoughtful discussions.  
We also thank the anonymous referee for constructive feedback. 
G.~R.~T.~is grateful  to R.~A.~S.,  and acknowledges support  from the
NASA/NY  Space  Grant  Consortium,  as  well as  a  European  Southern
Observatory  (ESO)   Fellowship  partially  funded   by  the  European
Community's Seventh Framework  Programme (/FP7/2007-2013/) under grant
agreement No.~229517.  Partial support was provided by NASA through an
award issued  by JPL/Caltech, as  well as the Radcliffe  Institute for
Advanced   Study  at   Harvard  University.    T.~E.~C.~was  partially
supported by NASA through {\it  Chandra} award G06-7115B issued by the
{\it Chandra} X-ray Observatory Center for and on behalf of NASA under
contract NAS8-39073. Basic research  into radio astronomy at the Naval
Research  Laboratory is  supported  by 6.1  Base funds.   C.~L.~S.~was
supported  in part  by NASA  {\it  Herschel} Grants  RSA 1373266,  RSA
P12-78175  and {\it Chandra}  Grant G01-12169X. A.~C.~F.~thanks the Royal Society.  This paper  is based
upon observations  with the {\it Chandra X-ray  Observatory}, which is
operated  by  the Smithsonian  Astrophysical  Observatory  for and  on
behalf  of NASA  under
contract NAS8-03060.   We also  make use of
previously published  observations by  the NASA/ESA {\it  Hubble Space
  Telescope}, obtained at the Space Telescope Science Institute, which
is  operated  by  the  Association  of Universities  for  Research  in
Astronomy, Inc., under NASA  contract 5-26555.  
The National Radio Astronomy Observatory is a facility of the 
National Science Foundation operated under cooperative agreement by 
Associated Universities, Inc.
We have made extensive
use of  the NASA Astrophysics  Data System bibliographic  services and
the NASA/IPAC  Extragalactic Database, operated by  the Jet Propulsion
Laboratory,  California Institute of  Technology, under  contract with
NASA.

\bibliographystyle{mn2e}
\bibliography{a2597_complete}

\label{lastpage}

\end{document}

%% file: journals.tex
%
\def\aj{AJ}%
\def\araa{ARA\&A}%
\def\apj{ApJ}%
\def\apjl{ApJ}%
\def\apjs{ApJS}%
\def\ao{Appl.~Opt.}%
\def\apss{Ap\&SS}%
\def\aap{A\&A}%
\def\aapr{A\&A~Rev.}%
\def\aaps{A\&AS}%
\def\azh{AZh}%
\def\baas{BAAS}%
\def\jrasc{JRASC}%
\def\memras{MmRAS}%
\def\mnras{MNRAS}%
\def\pra{Phys.~Rev.~A}%
\def\prb{Phys.~Rev.~B}%
\def\prc{Phys.~Rev.~C}%
\def\prd{Phys.~Rev.~D}%
\def\pre{Phys.~Rev.~E}%
\def\prl{Phys.~Rev.~Lett.}%
\def\pasp{PASP}%
\def\pasj{PASJ}%
\def\qjras{QJRAS}%
\def\skytel{S\&T}%
\def\solphys{Sol.~Phys.}%
\def\sovast{Soviet~Ast.}%
\def\ssr{Space~Sci.~Rev.}%
\def\zap{ZAp}%
\def\nat{Nature}%
\def\iaucirc{IAU~Circ.}%
\def\aplett{Astrophys.~Lett.}%
\def\apspr{Astrophys.~Space~Phys.~Res.}%
\def\bain{Bull.~Astron.~Inst.~Netherlands}%
\def\fcp{Fund.~Cosmic~Phys.}%
\def\gca{Geochim.~Cosmochim.~Acta}%
\def\grl{Geophys.~Res.~Lett.}%
\def\jcp{J.~Chem.~Phys.}%
\def\jgr{J.~Geophys.~Res.}%
\def\jqsrt{J.~Quant.~Spec.~Radiat.~Transf.}%
\def\memsai{Mem.~Soc.~Astron.~Italiana}%
\def\nphysa{Nucl.~Phys.~A}%
\def\physrep{Phys.~Rep.}%
\def\physscr{Phys.~Scr}%
\def\planss{Planet.~Space~Sci.}%
\def\procspie{Proc.~SPIE}%